\documentclass[11pt]{article}
\usepackage[margin=1in]{geometry}
\usepackage{amsthm,amsmath,natbib,amssymb,amsfonts}
\usepackage{bm}

\usepackage[section]{algorithm} 
\usepackage{algpseudocode}

\usepackage{graphicx}
\usepackage{subfig}
\usepackage{booktabs}
\usepackage{array}
\usepackage{paralist}
\usepackage{caption}
\usepackage{verbatim}
\usepackage{multirow}
\usepackage{booktabs}
\usepackage{rotating}
\usepackage{hyperref}
\usepackage{tikz}
\usepackage{subfig}
\usetikzlibrary{shapes,shadows,calc,arrows}
\usepgflibrary{arrows}

\bibpunct{(}{)}{;}{a}{,}{,}

\newcommand{\var}{\text{var}}
\newcommand{\cov}{\text{cov}}
\newcommand{\ESS}{\text{ESS}}

\newcommand{\Po}{\text{Poisson}}
\newcommand{\lN}{\text{logN}}

\newcommand{\vnorm}[1]{\left\vert#1\right\vert}
\algtext*{EndIf}
\algtext*{Until}
\newcommand{\bs}{}

\usepackage{dcolumn}
\newcolumntype{d}{D{.}{.}{-1}}

\title{RMCMC:\ A System for Updating Bayesian Models}

\author{F. Din-Houn Lau \qquad Axel Gandy \\Department of Mathematics,
  Imperial College London}

\date{}

\begin{document}
\maketitle

\begin{abstract}
  A system to update estimates from a sequence of probability
  distributions is presented. The aim of the system is to quickly
  produce estimates with a user-specified bound on the Monte Carlo
  error. The estimates are based upon weighted samples stored in a
  database. The stored samples are maintained such that the accuracy
  of the estimates and quality of the samples is satisfactory. This
  maintenance involves varying the number of samples in the database
  and updating their weights. New samples are generated, when
  required, by a Markov chain Monte Carlo algorithm. The system is
  demonstrated using a football league model that is used to predict
  the end of season table. Correctness of the estimates and their
  accuracy is shown in a simulation using a linear Gaussian model.
\end{abstract}

{\bf Key words:} Importance sampling; Markov chain Monte Carlo methods; Monte
Carlo techniques; Streaming data; Sports modelling 

\section{Introduction}\label{sec:intro}
We are interested in producing estimates from a sequence of
probability distributions. The aim is to quickly report these
estimates with a user-specified bound on the Monte Carlo error. We
assume that it is possible to use MCMC methods to draw samples from
the target distributions. For example, the sequence can be the
posterior distributions of parameters from a Bayesian model as
additional data becomes available, with the aim of reporting the
posterior means with the variance of the Monte Carlo error being less
than $0.01$. We present a general system that addresses this problem.

Our system involves saving the samples produced from the MCMC sampler
in a database. The samples are updated each time there is a change of
sample space. The update involves weighting or transiting the samples,
depending on whether the space sample changes or not. In order to
control the accuracy of the estimates, the samples in the database are
maintained. This maintenance involves increasing or decreasing the
number of samples in the database. This maintenance also involves
monitoring the quality of the samples using their effective sample
size. See Table \ref{tab:control_variates} for a summary of the
control variables. Another feature of our system is that the MCMC
sampler is paused whenever the estimate is accurate enough. The MCMC
sampler can later be resumed if a more accurate estimate is
required. Therefore, it may be the case that no new samples are
generated for some targets. Hence the system is efficient, as it
reuses samples and only generates new samples when necessary.

Our approach has similar steps to those used in sequential Monte Carlo
(SMC) methods \citep{smcm,liu2008monte}, such as an update (or
transition) step and re-weighting of the samples. Despite the
similarities, SMC methods are unable to achieve the desired aims
considered in this paper. Specifically, even though SMC methods are
able to produce estimates from a sequence of distributions, it is
unclear how to control the accuracy of this estimate without
restarting the whole procedure. For example, consider the simulations
in \cite{gordon} where the bootstrap particle filter, a particular SMC
method, is introduced. In these simulations the posterior mean is
reported with the interval between $2.5$ and $97.5$ percentile
points. As these percentile points are fixed, there is no way to
reduce the length of the interval. The only hope of reducing the
interval is to rerun the particle filter with more particles, although
there is no guarantee. This conflicts with the aim of reporting the
estimates quickly. In practice, most SMC methods are concerned with
models where only one observation is revealed at a time (see
simulations in e.g.\ \cite{kitagawa2014computational},
\cite{del2006sequential}, \cite{chopin2002sequential}). Our framework
allows for observations to be revealed in batches of varying sizes;
see the application presented in \S
\ref{sec:football}.

\begin{table}[tbp]
  \caption{Summary of control variables}
  \centering
  \begin{tabular}{lll}
    \toprule
    Control Variable & Measurement & Target Interval \\ 
    \midrule
    Accuracy of estimates ($A$) & Standard deviation of estimates    & $\left[\beta_1,\beta_2\right]$ \\ 
    Quality of samples ($Q$) &  $\text{Effective sample size of database}/N_{\text{MAX}}$& $\left[\gamma_1,\gamma_2\right]$ \\ 
    \bottomrule
  \end{tabular}
  \label{tab:control_variates}
\end{table}

A potential application of the system is monitoring the performance of
multiple hospitals where the data observed are patient records and
estimate of interest relates to the quality of patient care at each
hospital. Controlling the accuracy of this estimate may relate to
controlling the proportion of falsely inspected hospitals. Another
example of a realistic application of the system is a football league
model (see \S \ref{sec:football}) where the data revealed are the
match results and the estimate of interest is the end of season rank
league table. Controlling the estimated rank may be of interest to
sports pundits and gamblers. 

In \S \ref{sec:process_descriptions} we define, in detail, the setup
we are considering. We then describe the separate processes of the
system. We also describe how to combine the weighted samples to form
the estimate of interest. Then in \S \ref{sec:est_acc} we present a
modified batch mean approach that we use to compute the accuracy of
the estimate.  In \S \ref{sec:football} we investigate the performance
of the system using a model for a football league. For this
application, the aim is to provide quick and accurate predictions of
the end of season team ranks as football match results are
revealed. We examine the performance of the system as the size of data
received increases. Currently, there is no theoretical proof that the
proposed system is stable, however simulations verify correctness of
the reported accuracy and the estimate. We present such a simulation
in \S \ref{sec:kalman_compare}, where we apply the system using a
linear Gaussian model and simulated data.  We conclude in \S
\ref{sec:conclusion} with a discussion of potential future topics of
research.

\section{Description of the System}\label{sec:process_descriptions}
\subsection{Setup}\label{sec:setup}
We now describe the settings we consider and the necessary operations
required for our system to function. Let
$(S_n,\mathcal{S}_n,\pi_n)_{n\in\mathbb{N}}$ be a sequence of
probability spaces. We are interested in reporting $\pi_ng_n=\int
g_n(x)d\pi_n(x)$ where $g_n$ is a, possibly multivariate, random
variable on $(S_n,\mathcal{S}_n,\pi_n)$. In order to implement our
system, the following operations are required:\
\begin{enumerate}
\item \textit{MCMC}: For all $n\in\mathbb{N}$, generate samples from an MCMC with
  stationary distribution $\pi_n$.
\item \textit{Weighting Samples}: For all $k\in \overline{D}:=\{j: S_{j-1}=S_j \text{ and
  }\pi_j\ll\pi_{j-1} \}$, the Radon-Nikodym derivative $\frac{d\pi_j}{d\pi_{j-1}}$ can be evaluated.
\item \textit{Transiting Samples}: For all $k\in D$ $\exists f_k : S_{k-1}\times [0,1]\rightarrow S_k$ such that $\xi\sim \pi_{k-1}$, $U\sim U[0,1] \implies f_k(\xi,U)\sim\pi_k$.
\end{enumerate}
The weighting operation enables us to weight previously generated
samples according to the latest measure. In the case where the sample
space changes or the Radon-Nikodym derivative is not defined, the
transition operation allows us to map the samples to the latest
measure. If such a transition function, in operation 3, is
unavailable, then the following may be used instead.
\begin{enumerate}
\item[$3^\prime.$] \textit{Transiting Samples}: For all $k\in D$ $\exists f_k : S_{k-1}\times [0,1]\rightarrow S_k$.
\end{enumerate}
This alternative transition operation allows us to map the samples
into the latest sample space of interest. 

\subsection{Global Variables}\label{sec:sample database}
The samples produced by the RMCMC process (\S \ref{sec:RMCMC}) are
stored in a database. Each sample is recorded to the database with a
production date and an information cut-off. The production date is the
date and time the sample was written to the database from the RMCMC
process. The information cut-off refers to the measure the MCMC was
targeting when the sample was produced. Lastly, each sample will be
enter the database with weight $1$. The maximum number of samples
allowed in the database is $N_{\text{MAX}}$. In \S \ref{sec:control}
we explain how the control process varies $N_{\text{MAX}}$ over
time. Further, we shall refer to the current number of samples in the
database as $N$. The deletion process (\S \ref{sec:deletion}) ensures
that $N\leq N_{\text{MAX}}$ by sometimes removing samples from the
database. A summary of the systems global variables is provided in
Table \ref{tab:global_variables} along with their descriptions.

\begin{table}[tbp]
  \caption{Description of the global variables in the rolling MCMC system.}
  \centering
  \begin{tabular}{l p{5in}}
    \toprule
    Variable & Description \\ 
    \midrule
    \textsc{rmcmc\_on} & Indicator if RMCMC process is supposed to be producing new samples.\\
    $N_{\text{MAX}}$ & Maximum number of samples allowed in the database.\\
    $N$ & Number of samples contained in the database.\\
    $(\xi_i,w_i)_{i=1,\dots,N}$ & The samples and their corresponding weights in the database. \\
    \bottomrule
  \end{tabular}
  \label{tab:global_variables}
\end{table}

\subsection{RMCMC Process}\label{sec:RMCMC}
The RMCMC process, summarized in Algorithm \ref{code:rmcmc_process},
is an MCMC method that changes its target without the need to
restart. When the target of interest changes from $\pi_{j-1}$ to
$\pi_{j}$ so does the target of the MCMC. The Markov chain continues
from the latest sample, making a transition step using $f_j$ if there
is a change of sample space. This ensures the next MCMC is exploring
the correct space. We continue from this sample in the hope that the
Markov chain converges faster to the updated target distribution than
a randomly chosen starting point. To allow the Markov chain to move
toward the updated target distribution we use a burn-in period where
the first $B_0$ samples are discarded after the target changes. This
burn-in period will also weaken the dependence between samples from
different target distributions. As this MCMC method is never reset and
continues from the last generated sample we refer to it as a rolling
MCMC (RMCMC) process.

\begin{algorithm}[tbp]
\caption{RMCMC process}
\label{code:rmcmc_process}
\begin{algorithmic}[1]
  \Statex Parameters:\ MCMC algorithm, $B_0$.
  \Repeat{ indefinitely.}

  \If{target changes to $\pi_{j}$}
  \State Set $B=B_0$.
  \State Update target of MCMC to $\pi_{j}$.
  \If{$j\in D$}

  \State Set current position of MCMC to $f_j(\xi,U)$ where $U\sim U[0,1]$ and $\xi$
  is the latest sample generated.

  \EndIf

  \EndIf

  \If{\textsc{rmcmc\_on=true}}

  \State Perform MCMC step.

  \If{$B=0$} write sample to database with weight $1$.
  \Else $\text{ }$ $B\leftarrow B-1$.
  \EndIf

  \Else $\text{ }$sleep for some time.

   \EndIf

  \Until{}
\end{algorithmic}
\end{algorithm}

The RMCMC process is only active when new samples are required as it
can be paused and resumed by the control process (\S
\ref{sec:control}). If the process is paused for long periods, it may
be the case that no samples are produced for some targets. In
Algorithm \ref{code:rmcmc_process} the generated samples are written
to the database individually. In practice, however, it may be more
convenient to write the samples to the database in batches. This
practice is allowable and will not affect the functioning of the
system.

\subsection{Update Process}\label{sec:reweight}
The update process, presented in Algorithm
\ref{code:reweight_process}, ensures that the samples are weighted
correctly each time the target changes. There are two types of updates
depending on the measures and their sample spaces. More precisely,
consider a change of target from $\pi_{j-1}$ to $\pi_{j}$. If $j\in
D$, that is the sample spaces differ or the Radon-Nikodym $d
\pi_{j}/d\pi_{j-1}$ is not defined, then the function $f_k$ is used to
map the samples in the database onto the new space. On the other hand,
if $j\notin D$, the samples are first re-weighted according to $d
\pi_{j}/d\pi_{j-1}$, then scaled. We now discuss this re-weight and
scaling steps in more detail.

Suppose that the RMCMC process produces the samples
$\xi_1,\dots,\xi_m\sim \pi_{j-1}$ where $\pi_{j-1}$ is the target of
interest. Next, suppose the the target changes from $\pi_{j-1}$ to
$\pi_{j}$.  In order to use the samples from the previous measure,
$\pi_{j-1}$, for estimating $\pi_{j}g_j$, the weights are updated as
follows. For $i=1,\dots,m$ define the updated weight $W_i$ from $w_i$
as
\begin{equation*}
  W_i= w_i v_i\quad\text{where}\quad v_i\propto \frac{d\pi_{j}}{d\pi_{j-1}}(\xi_i).
\end{equation*}
After, the weights are scaled such that the sum of the weights is
equal to their effective sample size. More precisely, define the
scaled weight $\widehat{w}_i$ from $W_i $ as
\begin{equation*}
  \widehat{w}_i = W_i \frac{\sum_{k=1}^mW_k}{\sum_{k=1}^mW_k^2} \quad (i=1,\dots,m).
\end{equation*}
Straightforward calculations show that scaling in this fashion
ensures the effective sample size of the database is the sum of the
effective sample sizes of the most recently weighted samples and the
newly generated samples.

\begin{algorithm}[H]
  \caption{Update Process}
  \label{code:reweight_process}
  \begin{algorithmic}[1]
    \Repeat{ indefinitely.}
    \If{the target changes from $\pi_{j-1}$ to $\pi_{j}$}
    \State{Label the out-of-date samples $\xi_1,\dots\xi_m$ with corresponding weights $w_1,\dots,w_m$.}

    \If{$j\notin D$}
    \State Update the weight $w_i\leftarrow w_i v_i$ where $v_i\propto \frac{d\pi_{j}}{d\pi_{j-1}}(\xi_i)$ for $i=1,\dots,m$.
    \State Compute $d= \left(\sum_{k=1}^mw_k\right)/\left(\sum_{k=1}^mw_k^2\right)$.
    \State Set $w_i\leftarrow d w_i $ for $i=1,\dots,m$.
    \State Write the weights into the sample database.

    \EndIf

    \If{$j\in D$} 

    \State Replace samples by $f_j(\xi_i,U_1),\dots,f_j(\xi_m,U_m)$ where $U_1,\dots,U_m\stackrel{\text{iid}}{\sim} U[0,1]$, leaving the weights unchanged.

    \EndIf

    \Else
    $\text{ }$ sleep for some time.
    \EndIf
    \Until
  \end{algorithmic}
\end{algorithm}

\subsection{Control Process}\label{sec:control}
The control process determines when the RMCMC process is paused and
changes the maximum number of samples contained in the database. This
is done to maintain the accuracy of the estimate of interest and the
quality of the samples. We now discuss each of these in turn.

At any given time, denote the samples in the database by
$\xi_1,\dots,\xi_N$. For $i=1,\dots,N$ denote the $i$th sample weight
in the database as $w_i$. To estimate the quantity of interest
$\pi_k g_k$, for some $k\in \mathbb{N}$, we use the estimator
\begin{equation*}
  T=\frac{\sum_{i=1}^Nw_ig_k(\xi_i)}{\sum_{i=1}^Nw_i}.
\end{equation*}
The accuracy of the estimate, $A$, is defined as the standard
deviation of $T$ (in \S \ref{sec:est_acc} we discuss how to estimate
$A$). The process aims to control the accuracy $A$ such that
$A<\epsilon$ for some fixed $\epsilon>0$. When considering multiple
estimates i.e.\ multivariate $g_k$, we force the standard
deviation of all the estimates below the threshold $\epsilon$. One
approach to control the accuracy would be to pause the RMCMC process
each time $A<\epsilon$ and resumed if $A\geq\epsilon$. However, this
may lead to the RMCMC process being paused and resumed each time a new
observation is revealed, as a small change in the accuracy will
inevitably occur. Therefore, we use $0<\beta_1<\beta_2\leq\epsilon$ so
that if $A\leq \beta_1$ the RMCMC process is paused and if $A>\beta_2$
the RMCMC process is resumed.

The control process is also controls the quality of the samples
in the database. The process aims to hold a good mixture of samples in
the hope that a future change of measure does not require the resuming of the
RMCMC process. We define the quality of the samples in the database
as
\begin{equation*}
  Q=\frac{\text{ESS}}{N_{\text{MAX}}}\quad\text{where}\quad\text{ESS}=\frac{\left(\sum_{i=1}^Nw_i\right)^2}{\sum_{i=1}^Nw_i^2}.
\end{equation*}
The quality of the samples, $Q$, is the effective sample size of the
all the weights in the database divided by the optimal effective
sample size of the database, $N_{\text{MAX}}$. The optimal effective
size of the database consists of a database with $N_{\text{MAX}}$
samples all with weight 1. As with the accuracy, we aim to maintain
the quality such that $\gamma_1<Q<\gamma_2$ for some
$0<\gamma_1<\gamma_2\leq 1$.  The control process is summarised in
Algorithm \ref{code:control_process}. To ensure that the database in
never depleted, a minimum number of samples is imposed at
$N_{\text{MIN}}>0$ such that the number of samples, $N$ and
$N_{\text{MAX}}$ cannot drop below $N_{\text{MIN}}$. Therefore, when
the RMCMC process is paused, $Q<\gamma_1$ and
$N_{\text{MAX}}=N_{\text{MIN}}$ we cannot decrease $N_{\text{MAX}}$
any more. In this case, the RMCMC process is resumed to generate new
samples that replace the poor quality samples in the database.
\begin{algorithm}[btp]
  \caption{Control Process}\label{code:control_process}
  \begin{algorithmic}[1]
    \Statex{Parameters:\ $\beta_1,\beta_2,\gamma_1,\gamma_2,N_{\text{MIN}}$.}
    \Repeat{ indefinitely.}
    \State Compute $Q$ and $A$.
    \If{$A < \beta_1$ and $N\geq N_{\text{MIN}}$} 
    set \textsc{rmcmc\_on=false}.
    \EndIf
    \If{($A >\beta_2$) or (\textsc{rmcmc\_on=false} and $Q<\gamma_1$ and $N=N_{\text{MIN}}$)} set \textsc{rmcmc\_on=true}
    \EndIf
    \If{\textsc{rmcmc\_on=false} and $Q<\gamma_1$ and $N>N_{\text{MIN}}$}
     decrease $N_{\text{MAX}}$.
    \EndIf
    \If{\textsc{rmcmc\_on=true} and $Q>\gamma_2$}
     increase $N_{\text{MAX}}$.
    \EndIf
    \Until
  \end{algorithmic}
\end{algorithm}

\subsection{Deletion Process}\label{sec:deletion}
This process deletes samples from the database if the current number
of samples, $N$, exceeds the maximum number of samples allowed
$N_{\text{MAX}}$. Removing samples from the database reduces the
computational work performed by the update process and calculating
the estimates. Moreover, lowering the number of samples is the way the
control process maintains the quality of the samples. For simplicity,
if $N>N_{\text{MAX}}$, the $N-N_{\text{MAX}}$ samples that were
produced the earliest are removed.  The deletion process is summarized
in Algorithm \ref{code:deletion_process}.
\begin{algorithm}[btp]
  \caption{Deletion Process}\label{code:deletion_process}
  \begin{algorithmic}[1]
    \Repeat{ indefinitely.}
    \If{$N > N_{\text{MAX}}$} 
    delete samples from the database.
    \Else
    $\text{ }$ sleep for some time.
    \EndIf
    \Until
  \end{algorithmic}
\end{algorithm}

\subsection{Modifying Batch Means to Estimate the Accuracy}\label{sec:est_acc}
There are several methods to estimate the variance of MCMCs such as
block bootstrapping \citep[][Chapter 3]{lahiri_resampling}, batch
means \citep{flegal_batch} and initial sequence estimators
\citep{practical_MCMC_geyer}. In our system the samples in the
database have weights which complicates estimation of the
variance. The aforementioned methods cannot be used as they
essentially treat all samples with equal weight. We now present a
version of the batch mean approach that is modified to account for the
sample weights.

Assume the estimate of interest is $\pi_n g_n$ for some
$n\in\mathbb{N}$.  First, order the samples in the database
$\xi_1,\dots,\xi_N$ and their corresponding weights $w_1,\dots,w_N$ by
their production date. This ensures that the dependence structure of
the samples is maintained. Then we divide the samples into batches or
intervals of length $b$ according to their weights. More precisely,
let $D_0=0$, $D_j=\sum_{i=1}^jw_i$ and $L=\lceil \sum_{i=1}^Nw_i/b
\rceil$ be the number of batches. It may be the case that a weight
spans more than one interval. Therefore we need to divide each weight
by the proportion it spans a given interval. For the $i$th interval
and $u$th sample define $\kappa_i(u)=\left[ \min\left\{D_u,ib
  \right\}-\max\left\{D_{u-1},(i-1)b \right\} \right]^+$, where
$\left[ x \right]^+=\max\left( 0, x\right)$, for $i=1,\dots,L$. Then
$\kappa_i(u)$ is the batch weight of $\xi_u$ in interval $i$. The mean
of the weighted samples in the $i$th interval is
\begin{equation*}
  \widehat{\mu}_i=\frac{\sum_{u=1}^N\kappa_i(u)g_n(\xi_u)}{\sum_{u=1}^N\kappa_i(u)}.
\end{equation*}
Finally, we estimate the squared accuracy by
\begin{equation*}
  \hat{A}^2=\frac{1}{L} \sum_{i=1}^L \left(\widehat{\mu}_i-\widehat{\mu} \right)^2
   ,\quad \widehat{\mu}=\frac{1}{L}\sum_{j=1}^L\widehat{\mu}_j\text{  }.
\end{equation*}
The batch length $b$ should be large enough to capture the correlation
between samples, yet small enough to give a stable estimate of the
variance. In practice we recommend using several batch lengths in
order to get a conservative estimate of $A$. Moreover, the batch mean
estimate should not be trusted when the number of batches, $L$, is
low. This can occur as $\sum_{i=1}^Nw_i$ can become very small. In
this case, we suggest setting the accuracy $A$ to $-1$ nominally. This
prompts the control process to remove samples from the database and
then restart the RMCMC process. This action effectively replenishes
the database with new samples. In practice, we recommend taking this
action when $L<20$.

\subsection{Remarks}\label{sec:practical-notes}

\subsubsection{Effective sample size for correlated samples.}
The quality, $Q$, uses the effective sample size defined for
independent samples, not correlated samples which we use in our
system. In the system, consider the extreme case where all samples
have the same value i.e.\ $\xi_1=\dots=\xi_N$ produced from the same
target. Each of these samples will have the same weight and therefore
$Q=1$ suggesting the optimal quality has been achieved. Further, the
accuracy of the estimate, $A$, will be very low since the weights and
samples are all the same. Hence, in this extreme case, the control
process would take no action. This is clearly undesirable. Ideally,
the effective sample size used to calculate $Q$, should take into
account the autocorrelation of the samples, where high autocorrelation
(in absolute value) leads to a lower effective sample size. However,
we use the this version the effective sample size for independent
samples as it is quick and simple to compute.

\subsubsection{Degeneracy of the Sample Weights.}
We now discuss how the system handles two types of degeneracy of the
sample weights. The first is where a single sample in the database has
most of the total weight and all other samples have $0$ or nearly $0$
weight. If this were to occur, the effective sample size, and
therefore the quality, $Q$, will be very low. In this case, the
control process will remove samples from the database before resuming
the RMCMC process. The second is where all sample weights are $0$ or
nearly $0$. As a consequence, the sum of the weights,
$\sum_{i=1}^Nw_i$, will be very low. Recall that the batch mean
approach uses $L=\lceil \sum_{i=1}^Nw_i/b \rceil$ batches where $b$ is
the length of the batch. Further, if $L<20$ the control process
removes samples from the database and resumes the RMCMC
process. Therefore, in the case where $\sum_{i=1}^Nw_i$ drops to low,
the sample database in replenished. To summarise, the system does not
attempt to avoid these types of degeneracy, but to take remedial
action when it does occur.

\subsubsection{Burn-in Periods.}
In the RMCMC process, we perform a burn-in each time a change of
measure occurs. In some cases, however, it may not be necessary, as we
now discuss. Assume we have the samples $\xi_1,\dots,\xi_m\sim
\pi_{j-1}$. Next, consider a change of measure from $\pi_{j-1}$ to
$\pi_j$ such that $j\in D$. In this case, a burn-in period is
unnecessary as the new chain starts at a representative of $\pi_j$,
namely $f_j(\xi_m,U)\sim\pi_j$ where $U\sim U[0,1]$. On the other hand,
if either the samples $\xi_1,\dots,\xi_m$ are not from $\pi_{j-1}$ or
the transition function in operation $3^\prime$, but not operation
$3$, is available, then a burn-in period is required. In any case,
performing a burn-in is mostly harmless.

\subsubsection{Subsampling.}
The samples produced from a MCMC method are correlated. If the
correlation of the samples is high then a large number of samples are
required to achieved the desired accuracy of the estimate. As a
consequence, the update process and the calculation of the estimate
would take a long time. To alleviate this problem we use subsampling.

Use of subsampling within an MCMC method entails saving only some
samples produced. More precisely, with a subsampling size $k$, every
$k$th sample is saved and the rest discarded. To choose the
subsampling size $k$, we suggest performing a pre-initialisation run
of the MCMC on the initial set of data. One approach, that we use in
our implementation of the system, is to vary $k$ until
$\rho:=\varsigma^2/\var\left\{g_1(\xi_1)\right\}\approx 2$ where $
\varsigma^2=\var\left\{g_1(\xi_1)\right\}+2\sum_{j=1}^\infty\cov\left\{g_1(\xi_1),g_1(\xi_{1+j})\right\}$.
We found that setting $\rho\approx 2$ worked well in our
implementations of the system, however may not be appropriate in all
applications. In practice, a method such as initial sequence methods
\citep{practical_MCMC_geyer} or a batch mean approach \citep[][\S
1.10.1]{MonteCarloHandbook} can be used to estimate $\varsigma^2$. We
chose to use the batch mean approach in our system.

If the initial Markov chain $\xi_1,\xi_2,\dots$ is Harris recurrent
and stationary with invariant distribution $\pi$, then by the Markov
chain central limit theorem \citep[e.g.][]{jones2004markov}
\begin{equation*}
  \sqrt{n}\left\{\frac{1}{n}\sum_{i=1}^ng(\xi_i)- \int g(x)\pi(dx)\right\} \xrightarrow[]{d} N\left(0,\varsigma^2\right)\quad\text{as}\quad n\rightarrow\infty.
\end{equation*}
Thus $\varsigma^2$ is the asymptotic variance of the Markov
chain. Hence, by choosing $\rho\approx 2$, we obtain
\begin{equation*}
  2\sum_{j=1}^\infty\cov\left\{g(\xi_1),g(\xi_{1+j})\right\}\approx \var\left\{g(\xi_1)\right\}
\end{equation*}
i.e.\ the sum of all covariance terms contributes as much as
$\var\left\{g(\xi_1)\right\}$ to $\varsigma^2$. This way, the covariance between
the samples is prevented from getting to large relative to
$\var\left\{g(\xi_1)\right\}$.

\subsubsection{Choice of Scaling.}
As discussed in \S \ref{sec:RMCMC}, the database will consist of
weighted samples from different target distributions. In \S
\ref{sec:control} the weighted sample average, $T$, is used to
estimate $\pi_j g_j$ for some $j\in\mathbb{N}$. In this subsection we
show that, due to the scaling of the weights (\S \ref{sec:reweight}),
the variance of $T$ is minimised under certain assumptions. A similar
calculation can be found in \cite{gramacy}.

We begin by showing that $T$ can be decomposed according to two sets
of samples. Denote the invariant measure of the RMCMC process at a
given time instance as $\pi_j$ for some known
$j\in\mathbb{N}$. Further, label the samples produced from this MCMC
targeting $\pi_j$ as $\xi_{m+1},\dots,\xi_N$ for some
$m\in\{0,\dots,N\}$. The case $m=N$ corresponds to the situation when
no samples have been produced from $\pi_j$. Label the remaining sample
as $\xi_{1},\dots,\xi_m$. These samples will have already been
weighted and scaled in previous iterations.

The estimator, $T$, can be decomposed according
to the two sets of samples as
\begin{equation*}
  T = \frac{\sum_{i=1}^m\widehat{w}_ig_j(\xi_i)+\sum_{i=m+1}^Ng_j(\xi_i)}{\sum_{i=1}^m\widehat{w}_i+(N-m)}
\end{equation*}
as $w_j=1$ for $j=m+1,\dots,N$. In terms of the updated weights, $T$
can be written as $T = \alpha T_1+(1-\alpha) T_2$ where
$T_1=\sum_{i=1}^mW_ig_j(\xi_i)/ \sum_{i=1}^mW_i$ and
$T_2=\sum_{i=m+1}^Ng_j(\xi_i)/(N-m)$ are the individual estimators of
$\pi_jg_j$ given by the two sets of samples and
\begin{equation*}
  \alpha=\frac{\ESS_m}{\ESS_m+(N-m)}\quad\text{where}\quad\ESS_m=\frac{\left(\sum_{i=1}^mW_i\right)^2}{\sum_{i=1}^mW_i^2}.
\end{equation*}
The choice of the scaling performed in the update process (\S
\ref{sec:reweight}) led to this choice of $\alpha$. We now show that
this choice of $\alpha$, under certain assumptions, minimises the
variance of $T$. Assume $\phi\in\mathbb{R}$ is a constant. Then the
variance of the estimator $T=\phi T_1 +(1-\phi) T_2$ is
$\var(T)=\phi^2\var(T_1)+(1-\phi)^2\var(T_2)$ where we assume that
$T_1$ and $T_2$, or more specifically the two sets of samples
$\xi_1,\dots,\xi_m$ and $\xi_{m+1},\dots,\xi_{N}$, are
independent. The variances of the individual estimators are
\begin{equation*}
  \var(T_1)=\frac{\sigma^2}{\ESS_m}\quad\text{and}\quad \var(T_2)=\frac{\sigma^2}{N-m}
\end{equation*}
where we assume $\var\left\{g_j(\xi_i)\right\}=\sigma^2$, for
$i=1,\dots,N$ and that the weights are constants. Upon differentiating
we find that setting $\phi$ to $\ESS_m/\{ \ESS_m+(N-m)\}$ minimises
$\var(T)$ thus regaining $\alpha$. These assumptions are unrealistic
in our setting. However, this motivates the use of a burn-in period
within the RMCMC process after new data are observed. Although we can
not guarantee independence between the sets of samples, the burn-in
period at least weakens their dependence.

\section{Application to a Model of a Football
  League}\label{sec:football}
In this section we demonstrate how the system performs on a model of a
football league. The data we use are the English Premier League
results from $2005/06$ to $2012/13$ season. In a season, a team plays
all other teams twice. For each match played, a team receives points
based on the number of goals they and their opponent scores. If a team
scores more goals than their opponent they receive $3$ points. If a
team score the same number of goals as their opponent they receive $1$
point. If a team scores less goals than their opponent they receive
$0$ points. The rank of each team is determined by their total number
of points, where the team with the highest number of points is ranked
$1$st. A tie of ranks then determined by goal difference and then
number of goals scored.

We are interested in the probability of each rank position for all teams
at the end of a season. The aim is to estimate these rank
probabilities to a given accuracy. Thus, in this application we are
concerned about maintaining the accuracy of multiple predictions.

Throughout this section, we use the following notation. Let $\bs I_p$ be
the $p \times p$ identity matrix and $\bs 1_p$ be a vector of $1$s of
length $p$. Further, let $\text{N}(\bs\mu,\bs\Sigma)$ denote a
multivariate normal distribution with mean $\bs\mu$ and covariance
matrix $\bs\Sigma$. Denote the cardinality of a set $A$ by
$\vnorm{A}$. We shall reserve the index $t=1,\dots,T$ for reference to
seasons. Lastly, let $\lN(\mu,\sigma^2)$ denote a log-normal
distribution i.e.\ if $X\sim \text{N}(\mu,\sigma^2)$ then $\exp(X)\sim
\lN(\mu,\sigma^2)$.

We begin by presenting a model for football game outcomes. The model
we use is similar to that presented in
\cite{doi:10.1080/01621459.1998.10474084} and \cite{RSSC:RSSC065}.

\subsection{Football League Model}\label{sec:PL_model}
Consider a model with hidden Markov process $X_t$ ($t\in\mathbb{N}$),
observed process $Y_t$ ($t\in\mathbb{N}$) and parameter $\theta$. The
observation $Y_t$ contains all observations for state $X_t$. Denote
the $j$th observation of state $t$ as $Y_{j,t}$. Next define the $k$th
observation batch of state $t$ as $\widetilde{Y}_{k,t}$ for
$k=1,\dots,c_t$ for some $c_t\geq 1$. For instance, if the
observations are batched in groups of $10$, the $k$th batch of state
$t$ is $\widetilde{Y}_{k,t}=Y_{10k-9 ,t},\dots,Y_{10k,t}$. In this
application section, we are interested in the model
\begin{equation}\label{eq:batch_ssm}
  \begin{cases}
      p(x_t|x_{1:t-1},\theta)=p(x_t|x_{t-1},\theta)\\
      p(\widetilde{y}_{k,t}|\widetilde{y}_{1:(k-1),t},y_{1:(t-1)},x_{1:t},\theta)=p(\widetilde{y}_{k,t}|x_t,\theta) \\
      p(x_1|\theta), p(\theta)
  \end{cases}.
\end{equation}
where $\widetilde{y}_{1:0,t}$ is an empty observation batch introduced
for notational convenience. In this section, the sequence of target
distributions is defined as follows. Let
$\varpi_{k,t}=p(x_{1:t},\theta|\widetilde{y}_{1:k,t},y_{1:(t-1)})$
for $t=1,2,\dots$ and $k=0,\dots,c_t$. Then, we are interested in
the targets $\pi_n=\varpi_{\varphi_1(n),\varphi_2(n)}$ for
$n\in\mathbb{N}$ where
\begin{equation*}
  \varphi_2(n)=\max\left\{j\in\mathbb{N} : (n-1) \geq \sum_{i=1}^{j-1}(c_i+1) \right\} ,\quad \varphi_1(n)=n-1-\sum_{i=1}^{\varphi_2(n)-1}(c_i+1),
\end{equation*}
where we set $\sum_{i=1}^0(c_i+1)=0$. The transition steps occur at
$k\in D=\{n\in\mathbb{N}:\varphi_1(n)=0 \}$.  In this application, the
transition functions $f_k$ ($k\in D$) are dictated by the model namely
$p(x_t|x_{t-1},\theta)$ in \eqref{eq:batch_ssm}.

We now describe the states $X_t$, the observations $Y_t$ and the
parameter $\theta$ in this football application. Each team is assumed
to have a strength value (in $\mathbb{R}$) which remains constant
within a season. Let $U_t$ be the set of teams that play in season
$t$, $X_{i,t}$ be the strength of team $i$ in season $t$ and $\bs
X_t=(X_{i,t})_{i\in U_t}$.  To condense notation, for any set
$A\subset U_t$ define $\bs X_{A,t}:=(X_{i,t})_{i\in A}$ and form the
parameter vector
$\bs\theta=(\lambda_H,\lambda_A,\sigma_p,\sigma_s,\eta,\mu_p)$, which
we now define.

At the end of every season, some teams are relegated and new teams are
promoted to the league. Denote the set of promoted teams that begin
season $t$ by $W_t$ and let $V_t=U_t\backslash W_t$ be the set of
teams that remain in the league from season $t-1$ to $t$. The promoted
teams strengths are introduced such that $ \bs
X_{W_t,t}|(\bs\theta,\bs X_{t-1}=\bs x_{t-1})\sim \text{N}\left(\mu_p
  \bs 1_{\left\vert W_t \right\vert} ,\sigma_p^2\bs I_{\left\vert W_t
    \right\vert}\right)$.  Thus any previous history in the league is
not used for a promoted team. From season $t-1$ to $t$, the strengths
of the teams that were not relegated are evolved such that $ \bs
X_{V_t,t}|(\bs\theta,\bs X_{t-1}=\bs x_{t-1})\sim \text{N}\left(\eta
  \bs C_t \bs x_{V_t,t-1},\sigma_s^2\bs I_{\left\vert V_t
    \right\vert}\right)$.  Thus between seasons, the strengths of the
teams that are not relegated are centered around $0$ and expanded
($\eta>1$) or contracted ($\eta<1$). Next, consider a match, in season
$t$, between home team $j$ and away team $k$ ($j,k\in U_t$). We assume
the number of home $G_{j,H}^k$ and away goals $G_{k,A}^j$ is modelled
by $ G_{j,H}^k|(\bs\theta,\bs X_t) \sim \Po
\left(\lambda_H\exp\left\{x_{j,t}-x_{k,t}\right\}\right)$ and $
G_{k,A}^j|(\bs\theta,\bs X_t)\sim
\Po\left(\lambda_A\exp\left\{x_{k,t}-x_{j,t}\right\}\right) $
independently of each other. The parameters $\lambda_H$ and
$\lambda_A$ are strictly positive and pertain to the home and away
advantage (or disadvantage) which is assumed to be the same across all
teams and all seasons. More precisely, $\lambda_H$ ($\lambda_A$) is
the expected number of home (away) goals in a match between two teams
of equal strength.  Finally, denote the results of season $t$ by
$Y_t$; the number of home and away goals for all games in season $t$.
For this football application, the sample space is
$S_n=\mathbb{R}^{20\varphi_2(n)+2}\times(\mathbb{R}^+)^4$.

For the first season strengths, we use an improper flat prior. For the
home and away advantage we take respective Gamma distribution priors
of shapes $5$ and $2$ and scales $5$ and $1$. For $(\eta,\sigma_s)$
and $(\mu_p,\sigma_p)$ we take their Jeffreys priors. Jeffreys prior
was used for both $(\eta,\sigma_s)$ and $(\mu_p,\sigma_p)$ after
considering the amount of information available for each
parameter. For instance, if 10 seasons are considered, only 9
transitions between seasons are available for the likelihood of
$(\eta,\sigma_s)$. Thus, using an informative prior would greatly
influence the posterior distribution. This can also be argued for the
promotion parameters $(\mu_p,\sigma_p)$.

\subsection{The MCMC Step}\label{sec:mcmc-step}
For the MCMC step in the RMCMC process (Algorithm
\ref{code:rmcmc_process}), we use a Metropolis-Hasting algorithm
\citep{annealing2}, \citep{HASTINGS01041970}. In general, a different,
potentially more complex MCMC method can be used. However, the system
does not rely on the choice of MCMC method, and will work with a
simple sampler, as demonstrated in this application. We use
independent proposal densities for the separate parameters. Due to the
high dimension of the combine states and parameter, we choose to
implement block updates \citep[Section
21.3.2]{MonteCarloHandbook}. This entails proposing parts of the state
and parameter at any stage. The proposals densities used and the block
updating is summarized in Algorithm \ref{code:MH-alg}. In the
algorithm we propose a new strength of a single season $80$\% of the
time and part of the parameter $\theta$ the remaining $20$\%. This was
done so that exploration of the chain was mainly focused on the
states. The proposal densities parameters were determined by
consideration of the acceptance rate in a pre-initialization run of
the MCMC. Lastly, the samples were written into the database in
batches of $1,000$.
\begin{algorithm}[tbp]
  \caption{Block Proposals for Metropolis-Hasting Algorithm}\label{code:MH-alg}
  \begin{algorithmic}[1]
    \Statex Given $(x_{1:t},\lambda_H,\lambda_A,\eta,\sigma_s,\mu_p,\sigma_p)$.
    \State Generate $u\sim\text{Uniform}(0,1)$.
    \If{$u < 0.8$}
    \State Generate $v\sim\text{Uniform}\{1,\dots,t\}$.
    \State Propose $x_v^*|x_v\sim\text{N}(\bs x_v,0.0002 I_{\vnorm{U_v}})$  
    \EndIf
    \If{$u \geq 0.8$}
     Generate $w\sim\text{Uniform}\{1,\dots,4\}$.
    \If{$w=1$}
     propose $\lambda_H^*|\lambda_H\sim\lN(\log(\lambda_H),0.01^2)$.
    \EndIf
    \If{$w=2$}
     propose $\lambda_A^*|\lambda_A\sim\lN(\log(\lambda_A),0.01^2)$.
    \EndIf
    \If{$w=3$}
     propose $\eta^*|\eta\sim\text{N}(\eta,0.01)$ and $\sigma_s^*|\sigma_s\sim\lN(\log(\sigma_s),0.005)$.
    \EndIf
    \If{$w=4$} 
    propose $\mu_p^*|\mu_p\sim \text{N}(\mu_p,0.0002)$ and
    $\sigma_p^*|\sigma_p\sim\lN(\log(\sigma_p),0.002)$.
    \EndIf
    \EndIf
  \end{algorithmic}
\end{algorithm}

\subsection{League Predictions}\label{sec:league-predictions}
In \S \ref{sec:PL_model} we introduced a model for the team strengths
and the outcome of football matches, in terms of goals scored. In \S
\ref{sec:mcmc-step} we presented the MCMC method which produces
samples used to estimate the states and parameters of the model. We
now explain how these samples are used to predict the end of season
ranks of each team, which is our estimates of interest i.e.\ $\pi_n
g_n$.

For each sample, all games in a season are simulated once. Thus each
sample gives a predicted end of season rank table. The distribution
across these predicted rank tables gives the estimated probabilities
of the ranks of each team. This distribution is the posterior summary
of interest whose accuracy we aim to control.

\subsection{System Parameters}\label{sec:system-parameters}
As mentioned in \S \ref{sec:practical-notes}, we performed a
pre-initialization run using $10,000$ samples to determine the
subsampling size. Based on the results from the 2005/06 to the 2009/10
season, we found that a subsample size of $80$ gave $\varsigma\approx
2$. We used a burn-in period of $B_0=10,000$ within the RMCMC
process. Within the control process we use $\beta_1=0.01$ and
$\beta_2=0.0125$ for the accuracy thresholds and $\gamma_1=0.1$ and
$\gamma_2=0.75$ for the quality thresholds. Whenever the control
process demanded a change in $N_{\text{MAX}}$, it was increased or
decreased by $10$\% of its current value. Finally, we set
$N_{\text{MIN}}=1,000$.

As mentioned in \S \ref{sec:league-predictions}, our estimate consists
of rank probabilities for each team i.e.\ each team has estimated
probabilities for ending the season ranked
$1\text{st},\dots,20\text{th}$. The accuracy of each of the $400$ rank
probabilities is calculated using the method presented in \S
\ref{sec:est_acc} using use two batch lengths $b=10$ and $b=50$. The
maximum standard deviation is reported as the accuracy of the estimate
to be conservative.

\subsection{Results}\label{sec:results}
The system is initialized with the results from the 2005/06 to 2009/10
seasons of the English Premier League. Using the samples from this
initialisation, we proceeded with $3$ separate runs of the system. The
system itself remained unchanged in each of the runs, however, the way
the results for the next $2$ seasons were revealed varied. The match
results were revealed individually, in batches of $7$ days and in
batches of $30$ days. New data batch were revealed only if the RMCMC
process was paused.

In Table \ref{tab:batch_sizes} we present the system results of each
run. We see that for larger data batches, the RMCMC process is resumed
more often. Further, the percentage of new samples generated after new
data are revealed increases as with the size of the data batch. The
average percentage of new samples is calculated as follows. Before a
new data batch is revealed the percentage of new samples in the
database generated after the introduction of the latest data is
calculated. The average of these percentages is then taken over the
data batches. This means that for larger data batches the RMCMC
process will often be resumed to generate new samples that replace
most of the samples already in the database.
\begin{table}[tbp]
  \caption{System summary for various data batch sizes.}
  \label{tab:batch_sizes}
  \centering
  \begin{tabular}{llll}
    \toprule
    & individual & 7 day & 30 day \\ 
    \midrule
    No. of batches & 760 & 70 & 20 \\ 
    Range of games per batch & [1,1] & [3,21] & [10,53] \\ 
    No. times RMCMC resumed & 39 & 31 & 18 \\ 
    Total No. MCMC steps & 24,010,000 & 14,230,000 & 9,240,000 \\ 
    Average \% of new samples & 2\% & 20\% & 53.6\% \\ 
    \bottomrule
  \end{tabular}
\end{table}
In Table \ref{tab:parameter_estimates} we present the estimated
posterior mean of the components of $\theta$ at the end of the run for
each batch size. As expected, being based on the same data, these
final estimates are almost identical for the various batch sizes. In
Table \ref{tab:condensed_rank_tab} we present the predicted end of
2012/13 season ranks for selected teams and ranks. Each team and rank
has $3$ predictions given by the runs using different batch sizes. For
each batch size, these predictions are being controlled. More
precisely, for every rank of every team the predictions standard
deviation is being controlled below $\beta_2=0.0125$. This is
consistent with the predictions across the various batch sizes. The
predictions for all teams and ranks can be found in \S \ref{sec:more
  results} in the Appendix.
\begin{table}[tbp]
\caption{End of run parameter estimated mean with $95$\% credible intervals.}
\centering
\begin{tabular}{clll}
  \toprule
 Parameter& individual & 7 day & 30 day \\ 
  \midrule
  $\lambda_H$ & 1.446 (1.406,1.497) & 1.447 (1.406,1.496) & 1.446 (1.406,1.494) \\ 
  $\lambda_A$ & 1.031 (0.998,1.073) & 1.032 (0.995,1.073) & 1.032 (0.997,1.077) \\ 
  $\eta$ & 0.970 (0.865,1.054) & 0.967 (0.865,1.049) & 0.964 (0.864,1.048) \\ 
  $\sigma_s$ & 0.083 (0.061,0.117) & 0.084 (0.059,0.113) & 0.086 (0.059,0.116) \\ 
  $-\mu_p$ & 0.245 (0.316,0.172) & 0.242 (0.322,0.167) & 0.244 (0.315,0.171) \\ 
  $\sigma_p$ & 0.116 (0.049,0.204) & 0.117 (0.063,0.191) & 0.114 (0.06,0.202) \\ 
  \toprule
\end{tabular}
\label{tab:parameter_estimates}
\end{table}
\begin{table}[tbp]
  \caption{End of 2012/13 season rank predictions for selected teams and ranks. Each team and rank has $3$ predictions given by (from top to bottom) the individual, $7$ day and $30$ day batch run.}
  \label{tab:condensed_rank_tab}
  \centering
  \begin{tabular}{lllllll}
    \toprule
    &\multicolumn{6}{c}{Rank}\\
    \cmidrule(r){2-7}
    Team & 1 & 2 & 3 & $\dots$ & 18 & 19 \\ 
  \midrule
  \multirow{3}{*}{Arsenal} & 8\% & 14\% & 17\% & $\dots$& 0\% & 0\% \\ 
  & 8\% & 14\% & 17\%& $\dots$ & 0\% & 0\% \\
  &8\% & 15\% & 19\%& $\dots$ & 0\% & 0\% \\
  \midrule
  \multirow{3}{*}{Aston Villa} & 0\% & 0\% & 1\%& $\dots$ & 6\% & 5\% \\ 
   & 0\% & 0\% & 1\%& $\dots$ & 6\% & 5\% \\ 
   & 0\% & 0\% & 1\%& $\dots$ & 6\% & 5\% \\ 
  \midrule
  \multirow{3}{*}{Chelsea} & 9\% & 15\% & 19\%& $\dots$ & 0\% & 0\% \\ 
   & 9\% & 15\% & 21\%& $\dots$ & 0\% & 0\% \\ 
   & 10\% & 16\% & 20\%& $\dots$ & 0\% & 0\% \\ 
  \midrule
  \multirow{3}{*}{Everton} & 1\% & 2\% & 6\%& $\dots$ & 1\% & 1\% \\ 
   & 1\% & 3\% & 5\%& $\dots$ & 1\% & 1\% \\ 
   & 1\% & 2\% & 5\%& $\dots$ & 1\% & 1\% \\ 
  \midrule
   \qquad \vdots & \vdots & \vdots & \vdots & \vdots  & \vdots & \vdots \\ 
  \midrule
  \multirow{3}{*}{Wigan} & 0\% & 0\% & 0\%& $\dots$ & 10\% & 11\% \\ 
   & 0\% & 0\% & 0\%& $\dots$ & 10\% & 11\% \\ 
   & 0\% & 0\% & 0\%& $\dots$ & 11\% & 12\% \\ 
   \bottomrule
\end{tabular}
\end{table}
In the following, we present some results for the $7$ day batch run
only. Further results for all the batch sizes are presented in \S
\ref{sec:more results} in the Appendix. In Figure
\ref{fig:system_res_weekly} we display the accuracy of the predictions
($A$), the quality of the samples ($Q$) and the number of samples in
the database ($N$) as new data are revealed. In Figure
\ref{fig:here_weekly_A}, the control process attempts to keep the
accuracy of the predictions between $\beta_1=0.01$ and
$\beta_2=0.0125$. Occasionally, after new data are revealed, the
accuracy exceeds the upper threshold $\beta_1$. The accuracy drops
nominally to $0$ at the end of each season prior to the introduction
of the next seasons fixtures. Similarly, in Figure
\ref{fig:here_weekly_Q}, the quality of the samples is attempted to be
kept between $\gamma_1=0.1$ and $\gamma_2=0.75$. In Figure
\ref{fig:here_weekly_N}, we see that the number of samples in the
database, $N$, varies over time. More precisely, after $5$ batches of
data, $19,246$ samples are used. However, later the number of samples
used decreases to approximate $14,000$ samples. Similar features are
seen for the different batch sizes. The change in the accuracy of the
predictions and the quality of the samples gets smaller as the batch
size decreases.

Figure \ref{fig:KM_bulk} is a plot of the Kaplan-Meier estimator
\citep{km} of the survival function of the samples in the database as
new data are revealed. More precisely, let $U$ be a random variable of
the number of new data batches observed before a sample is
deleted. Then Fig \ref{fig:KM_bulk} is a plot of the Kaplan-Meier
estimator of $S(u)=P(U>u)$. The Kaplan-Meier estimator takes into
consideration the right-censoring due to the end of the simulation
i.e.\ samples that could have survived longer after the simulation
ended. We see that samples survive as new data are observed e.g.\ a
sample survives $10$ or more batches with probability $0.33$. Thus
samples are reused as envisaged in \S \ref{sec:reweight}. Lastly, from
using the different batch sizes (see \S \ref{sec:more results} in the
Appendix) we see that samples survive more data batches as the size of
the batch gets smaller.

\begin{figure}[tbp]
  \centering
  \subfloat[]{\label{fig:here_weekly_A}\includegraphics[width=0.5\textwidth]{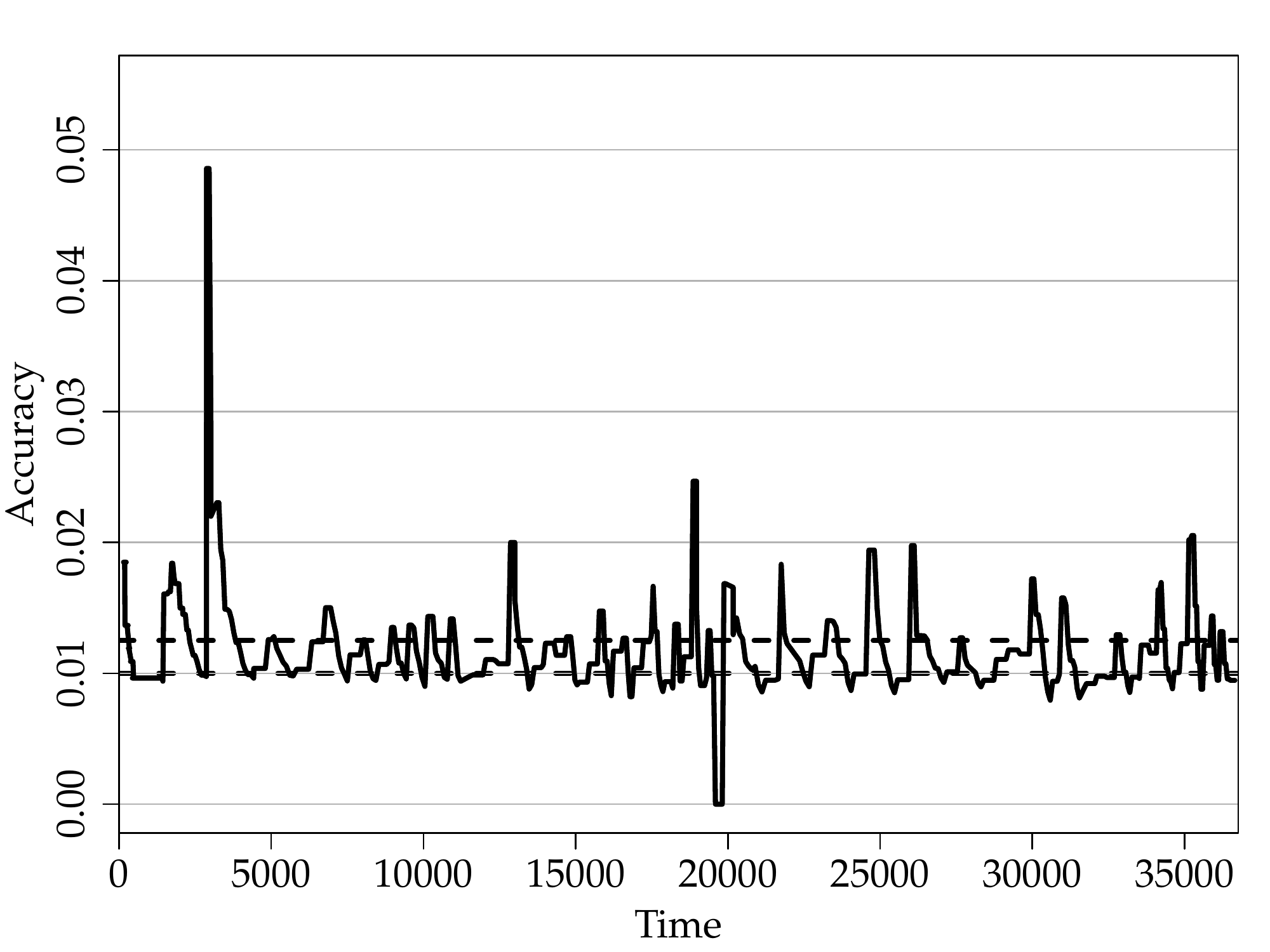}}
  \subfloat[]{\label{fig:here_weekly_Q}\includegraphics[width=0.5\textwidth]{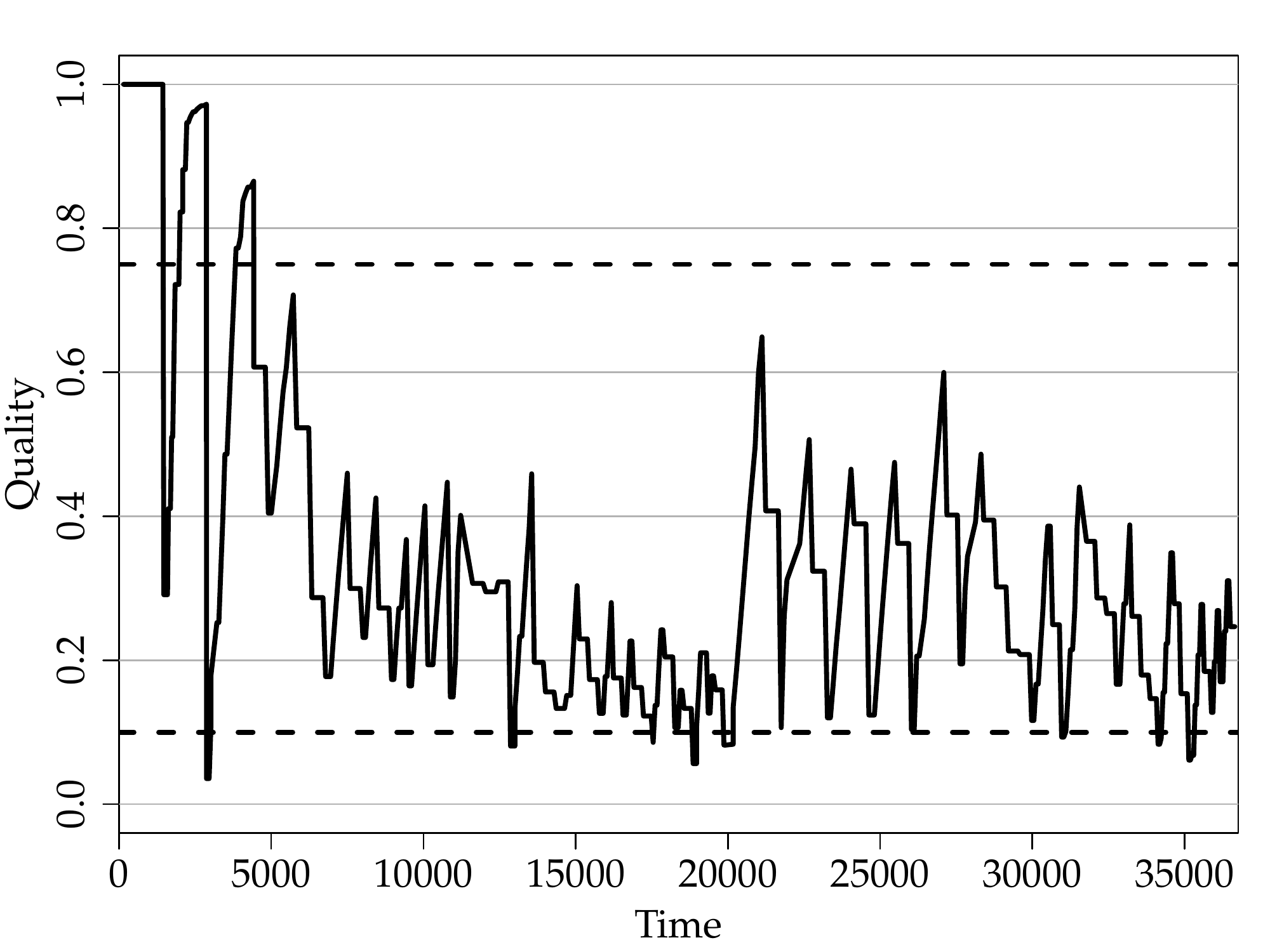}}%

  \subfloat[]{\label{fig:here_weekly_N}\includegraphics[width=0.5\textwidth]{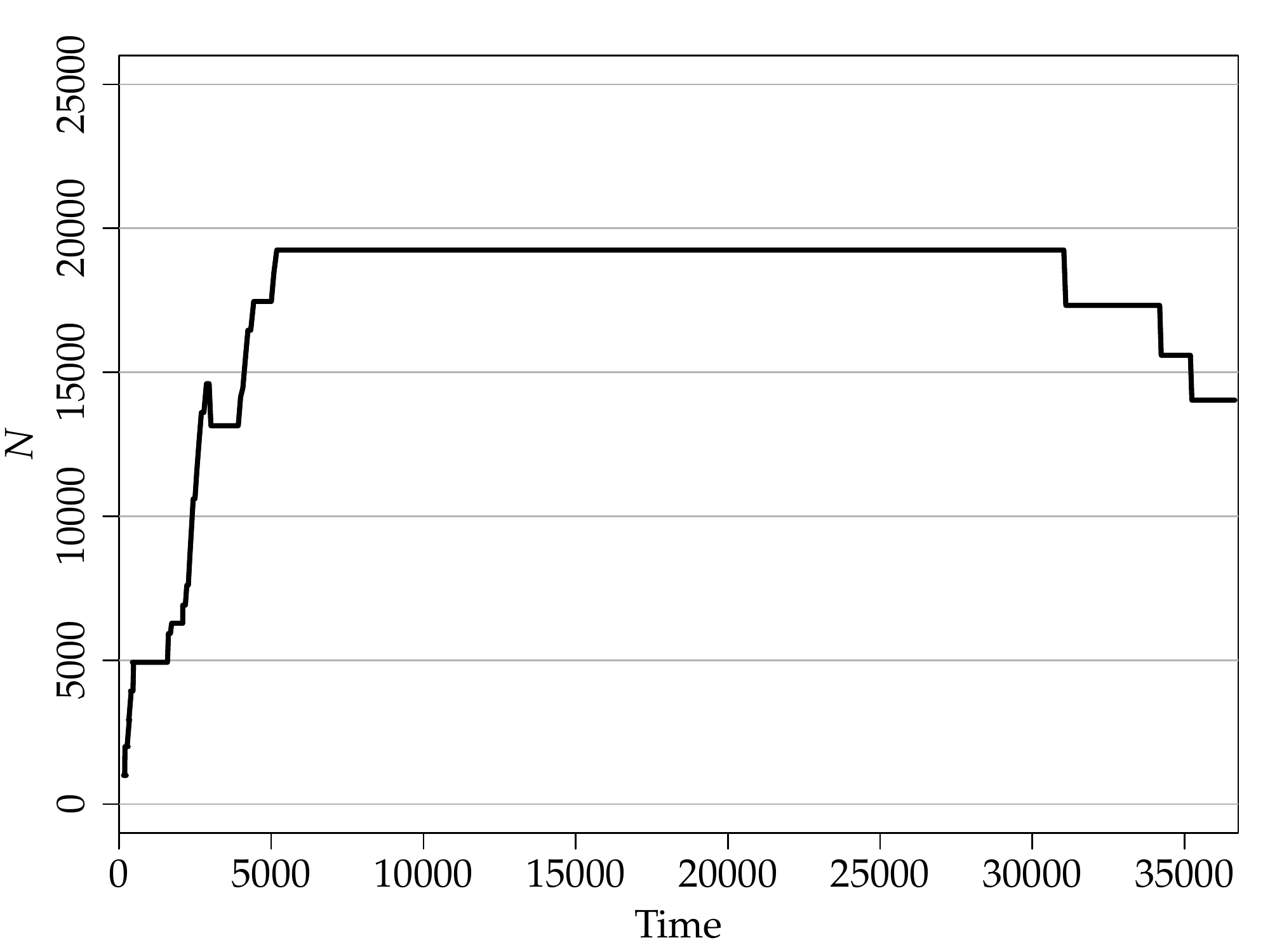}}
  \subfloat[]{\label{fig:KM_bulk}\includegraphics[width=0.5\textwidth]{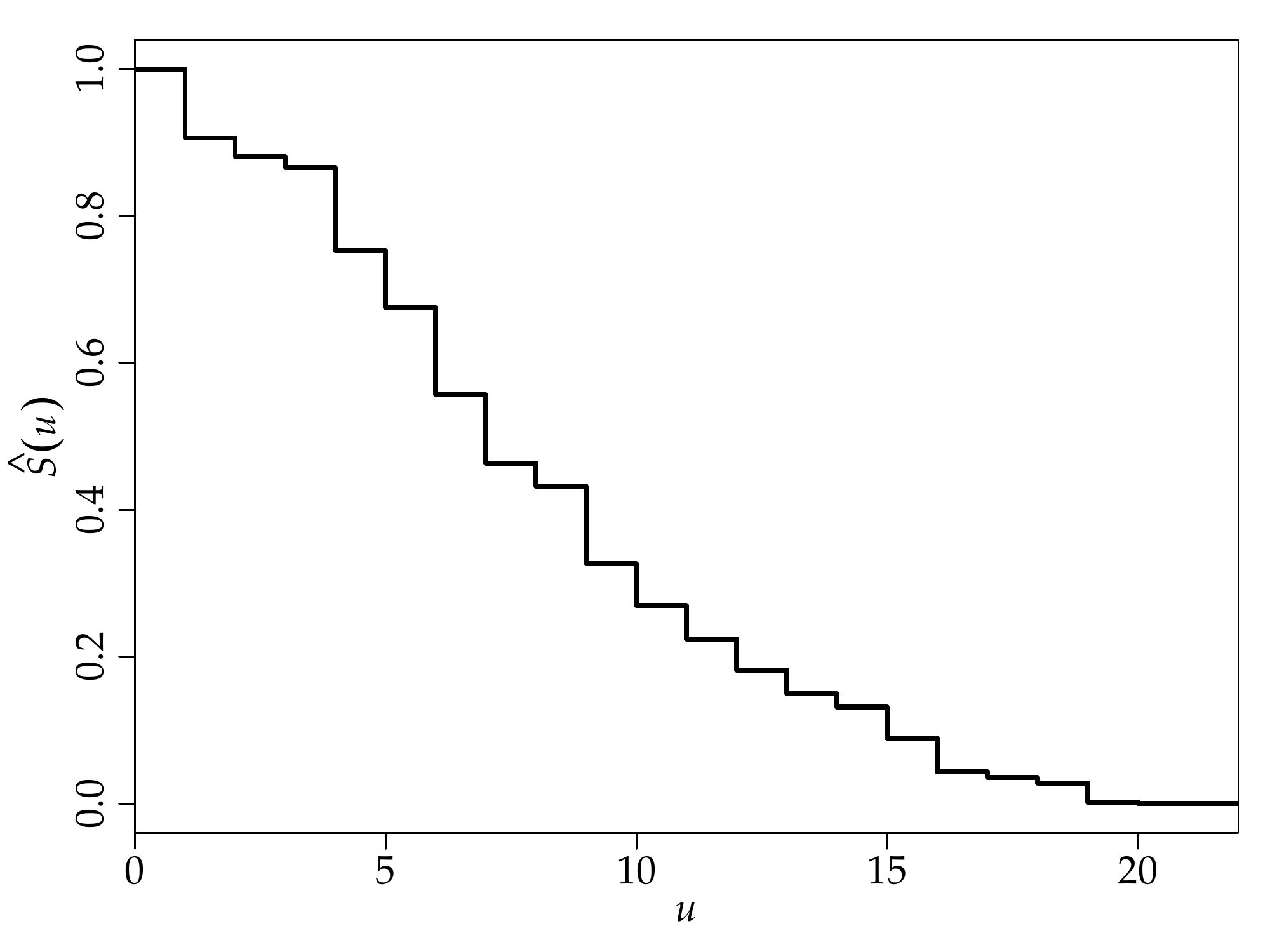}}
  \caption{System results using $7$ day batches where (a) is the
    accuracy ($A$) of the predictions, (b) is the quality ($Q$) of the
    samples, (c) is the number of samples in the database ($N$) as new
    data are revealed and (d) is the Kaplan-Meier estimator the
    samples lifetime.}
  \label{fig:system_res_weekly}
\end{figure}
In order to determine the quality of the predicted ranks given by the
system, we performed a separate run and consider the coverages of the
prediction intervals. For this run the initial observations consisted
of the 2005/06 to 2007/08 seasons results. We then introduced the
match results for the 2008/09 to 2012/13 seasons in $7$ day
batches. Over the $5$ seasons there were $178$ batches of intervals
for each team. Before each batch of results was introduced,
conservative $50$\% and $95$\% intervals were formed for the predicted
end of season rank of each team. These confidence intervals are
conservative due to the discreteness of ranks. The true mass contained
in the conservative $50$\% intervals was on average
$76.1$\%. Similarly, the true mass contained in the conservative
$95$\% intervals was on average $98.8$\%.  When compared with the true
end of season ranks, $74.2$\% of the true ranks lied in the
conservative $50$\% intervals and $99.3$\% lied in the $95$\%
intervals.

\section{Application to a Linear Gaussian Model}\label{sec:kalman_compare}
In \S \ref{sec:football} we are unable to check if the strengths of
the teams and the other parameters (i.e.\ the states and parameters)
are being estimated accurately as their true distributions are
unknown. In this section we inspect the estimates given by the RMCMC
system using simulated data. We use a linear Gaussian model such that
the Kalman filter \citep{kalman1960new} can be applied. This
simulation will allow us to compare the RMCMC system and the Kalman
filter estimates. This linear and Gaussian
model was chosen to resemble the football model described in \S
\ref{sec:football}. For this model the Kalman filter gives the exact
conditional distribution. Therefore, the Kalman filter will provide
the benchmark estimates to compare against.

Consider the model defined as follows:\
\begin{equation}
  \label{eq:lgm}
  \begin{cases}
    \text{State}:\quad\quad\quad X_t=A X_{t-1}+\Phi_t, & \Phi_t\stackrel{\text{iid}}{\sim}\text{N}(0,\Sigma)\\
    \text{Observation}:\quad Y_t=B X_{t}+\Psi_t, & \Psi_t\stackrel{\text{iid}}{\sim}\text{N}(0,\Xi)
  \end{cases},\quad\text{for } t=1,2,\dots
\end{equation}
and prior distribution $X_0\sim \text{N}(\mu_0,\Sigma_0)$. For this
particular simulation we chose $A=0.7
(\mathbb{I}_{20}-\frac{1}{20}1_{20}1_{20}^T)$,
$\Sigma=0.05\mathbb{I}_{20}$ and $\Xi=0.02 \mathbb{I}_{380}$. The
matrix $B$ is constructed according to the football matches in the
English Premier League in the $2005/06$ season. More precisely, each
row of $B$ is consists of zeros apart from two entries at $i$ and $j$
corresponding to a football match between home team $i$ and away team
$j$. A $2$ is put in the $i$th position and a $1$ at the $j$th. The
rows are ordered chronologically from top to bottom. For the prior
distribution, we set $\mu_0$ to be a vector of zeroes and
$\Sigma_0=\mathbb{I}_{20}$. Denote the $i$th component of $X_t$ as
$X_{i,t}$.

A single realisation of the states and observations were generated for
$t=1,\dots,7$. Using these observations, the RMCMC system was run
$100$ times to estimate means. This was compared with the estimates
given by the Kalman filer. Each run of the RMCMC system was
initialised using the observations from state $t=1,\dots,5$.

The sequence of targets is similar to that used in \S
\ref{sec:PL_model} with
$\varpi_{k,t}=p(x_{1:t}|\widetilde{y}_{1:k,t},y_{1:(t-1)})$. For the
transition function $f_j$ ($j\in D$) we use the observation equation
in \eqref{eq:lgm}. Finally, we take $g_n$ to be the identity function,
so that our estimate of interest in the posterior mean. The
observations were revealed in batches of $10$, so that each state
consisted of $38$ batches. Specifically, the vector $Y_t$ contains all
$380$ observations where we denote the $j$th observation as
$Y_{j,t}$. The $k$th observation batch of state $t$ is
$\widetilde{Y}_{k , t}:=Y_{10k-9,t},\dots,Y_{10k,t}$. Therefore, after
initialisation, the batches $\widetilde{Y}_{1 ,
  6},\dots,\widetilde{Y}_{38 , 6},\widetilde{Y}_{1 , 7},\dots
\widetilde{Y}_{38 , 7}$ are revealed.

Within the control process we again use $\beta_1=0.01$,
$\beta_2=0.0125$ and $\gamma_1=0.1$, $\gamma_2=0.75$. Also, we set
$N_{\text{MIN}}=1,000$. For this simulation controlling the accuracy
$A$ pertains to controlling the mean posterior of each component of
every state as new data are observed. We use a Gibbs sampler
\citep[see e.g.][]{geman1984stochastic} as the conditional
distributions for the states can be explicitly computed for this
model. Each Gibbs sampler step consists of updating a single randomly
chosen state as outlined in Algorithm \ref{code:gibbs-sampler}. We
used no subsampling and a burn-in period of $B_0=1,000$. The accuracy
was calculated using the batch mean approach described in \S
\ref{sec:est_acc} with batch lengths $10$ and $25$. The RMCMC process
wrote $500$ samples to the database at a time.
\begin{algorithm}[tbp]
  \caption{Gibbs Sampler:\ Single step}\label{code:gibbs-sampler}
  \begin{algorithmic}[1]
    \Statex Given $X_1,\dots,X_n$ and $Y_1,\dots,Y_n$.
    \State Generate $s\sim\text{Uniform}\{1,\dots,n\}$.
    \If{$s = 1$} Draw $Z_s$ from the pdf $f(x_1|X_2,\dots,X_n,Y_1,\dots,Y_n)$.
     \EndIf
     \If{$s >1$ and $s < n$} Draw $Z_s$ from the pdf $f(x_s|X_1,\dots,X_{s-1},X_{s+1},\dots,X_n,Y_1,\dots,Y_n)$.
    \EndIf
    \If{$s=n$}    Draw $Z_s$ from the pdf $f(x_s|X_1,\dots,X_{n-1},Y_1,\dots,Y_n)$.
    \EndIf
    \State Let $X_s=Z_s$.
  \end{algorithmic}
\end{algorithm}

\begin{figure}[tbp]
  \centering
  \subfloat[Posterior Mean of $X_{5,6}$ and $X_{5,7}$.]{\label{fig:team5}\includegraphics[width=0.5\textwidth]{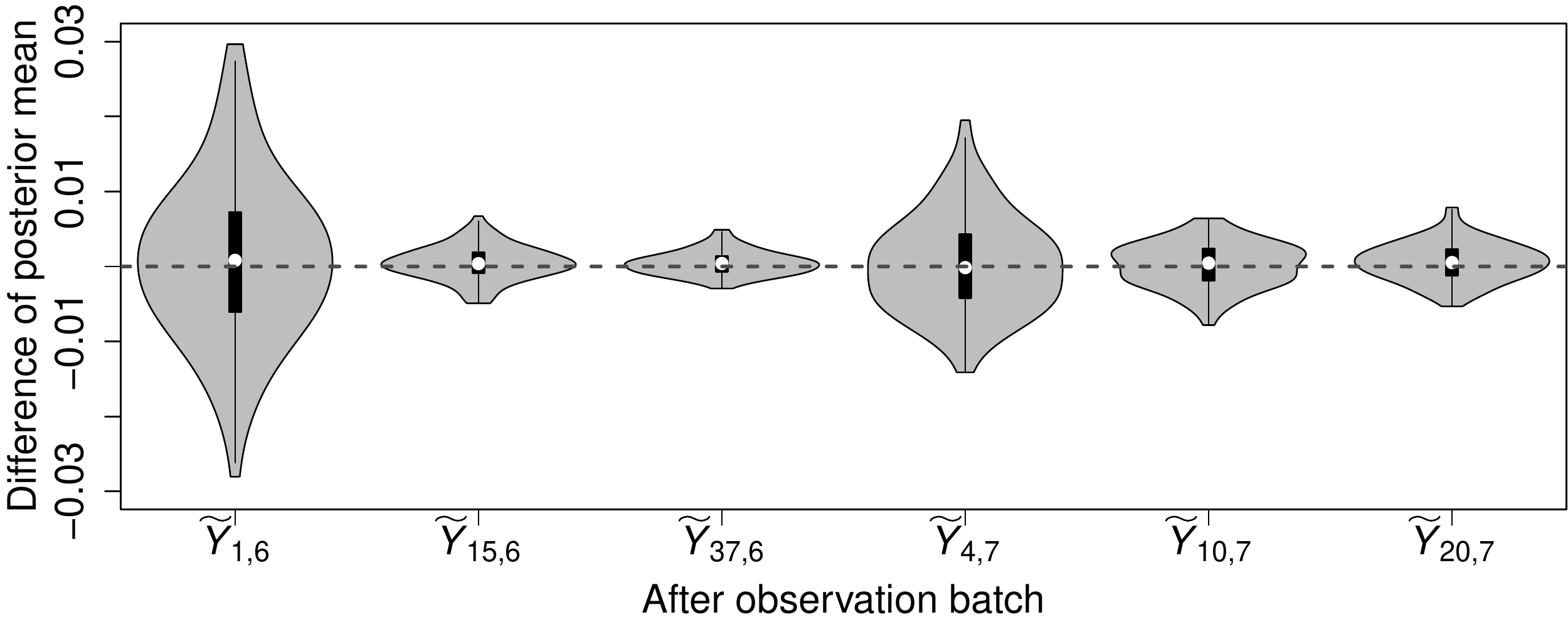}}
  \subfloat[Posterior Mean of $X_{18,6}$ and $X_{18,7}$.]{\label{fig:team18}\includegraphics[width=0.5\textwidth]{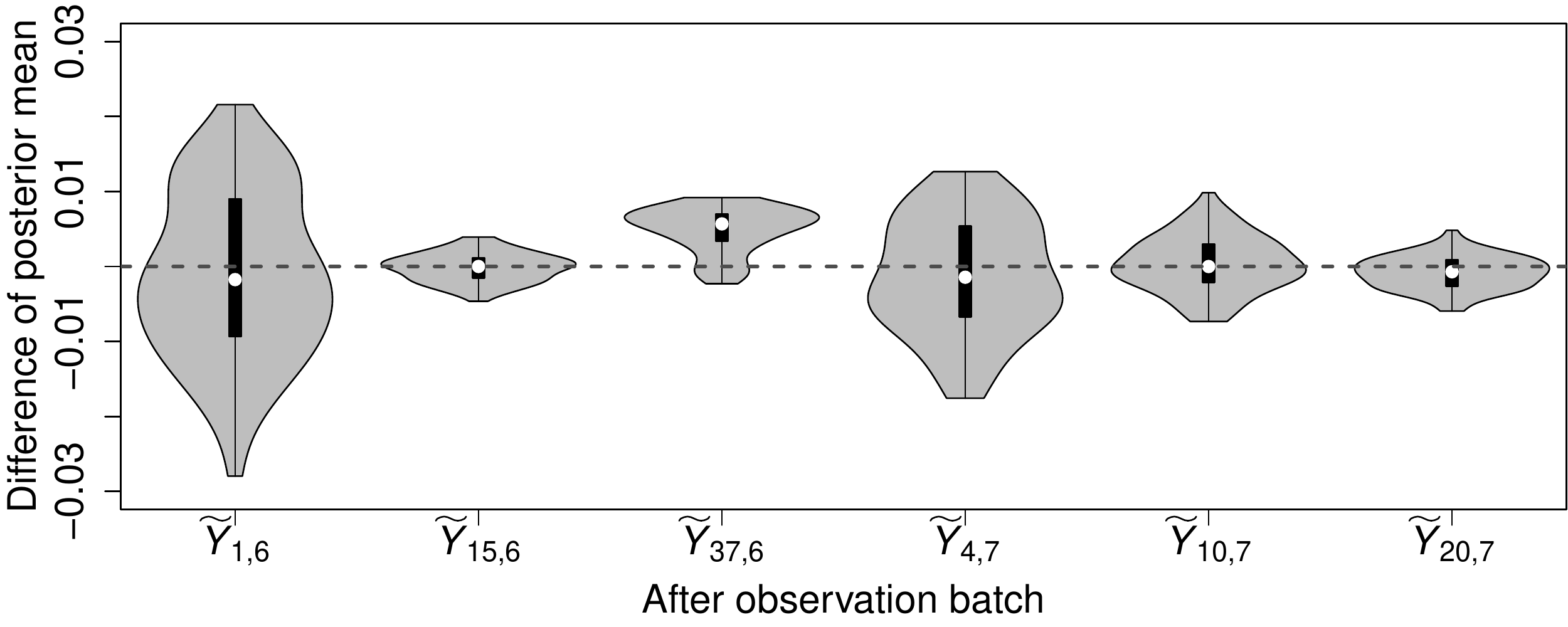}}

  \subfloat[Posterior mean of $X_{5,6}$ after $\widetilde{Y}_{1,6}$ revealed.]{\label{fig:QQ_1}\includegraphics[width=0.5\textwidth]{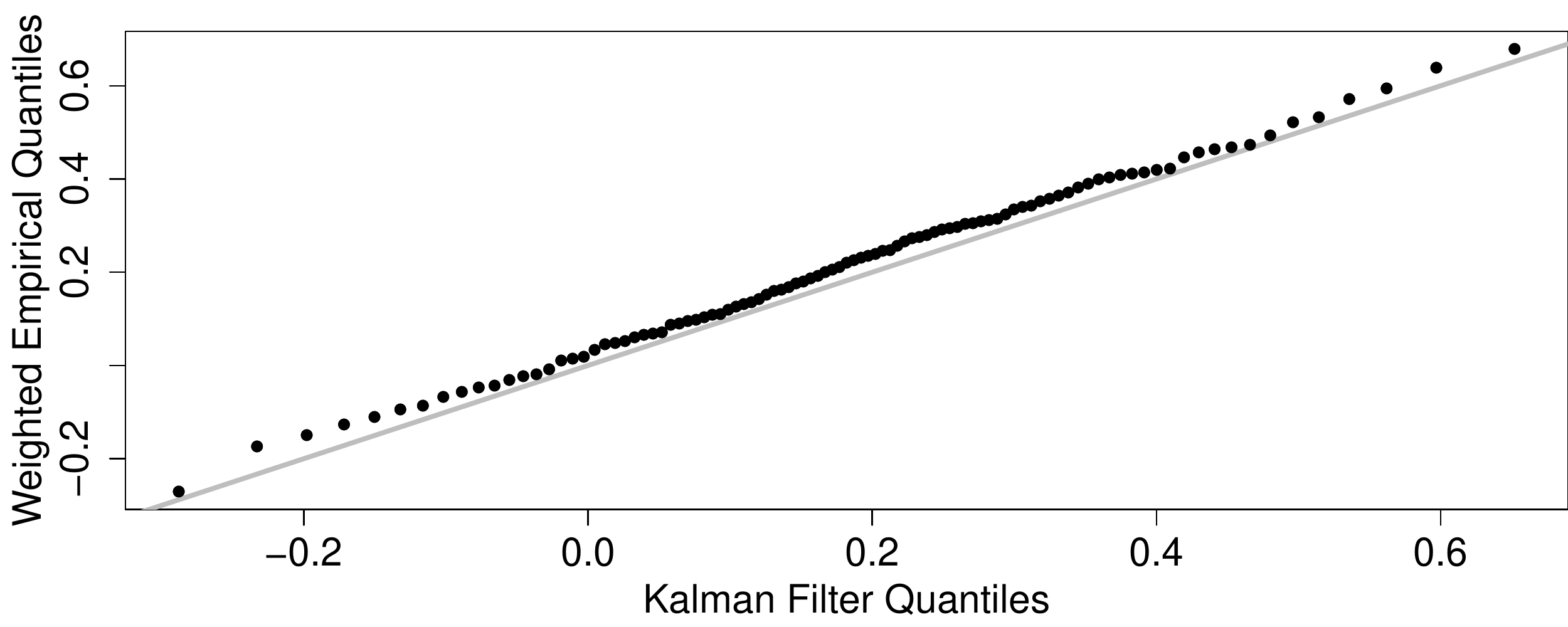}}
  \subfloat[Posterior mean of $X_{18,7}$ after $\widetilde{Y}_{20,7}$ revealed.]{\label{fig:QQ_2}\includegraphics[width=0.5\textwidth]{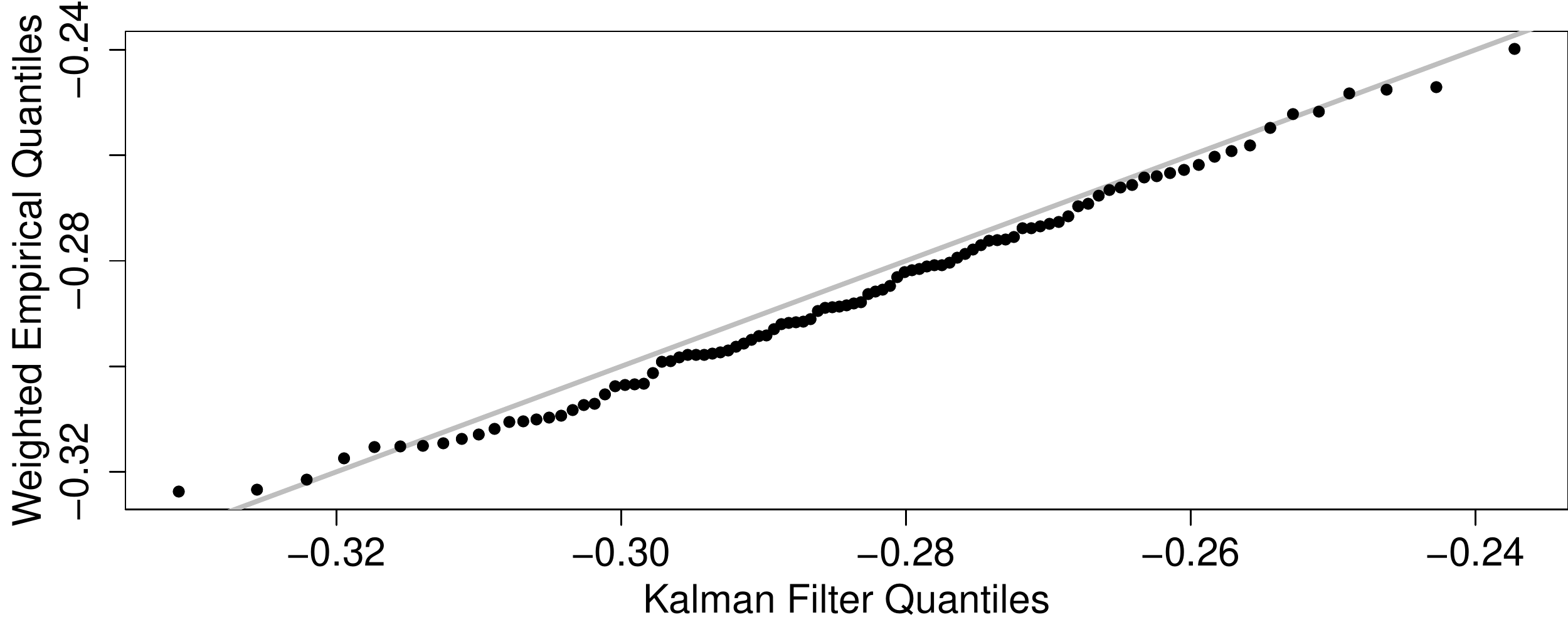}}
  \caption{Simulation results:\ (a) and (b) are violin plots of the
    difference between the $100$ RMCMC and the Kalman filter posterior
    mean. (c) and (d) are Q-Q plots of the Kalman filter and RMCMC
    system posterior distribution from a single run (black).}
  \label{fig:kf-plots}
\end{figure}

Figure \ref{fig:kf-plots} presents results comparing the Kalman filter
estimates with the $100$ RMCMC estimates as the observations are
revealed. The upper row of Fig \ref{fig:kf-plots} are violin plots
\citep[see e.g.\ ][]{violin-plots} of the difference between the
Kalman filter and the $100$ RMCMC system posterior mean of selected
states and components. Violin plots are smoothed histograms either
side of a box plot of the data.

The estimate may be bias due to the scaling and normalisation of the
weights carried out by the update process (\S \ref{sec:reweight})
(see for example \cite{hesterberg1995weighted} for the bias in
weighted importance sampling). This is apparent in posterior mean for
$X_{18,6}$ (Fig.~\ref{fig:team18}), as in $81$ out of the $100$ runs
the RMCMC process remained paused after $\widetilde{Y}_{37,6}$ was
revealed. For these $81$ runs, the posterior mean was formed using
weighted importance sampling. In contrast, we see nearly no bias in
the posterior mean for $X_{5,6}$ after $\widetilde{Y}_{1,6}$ was
revealed (Fig.~\ref{fig:team5}) where the RMCMC process was started in
every run (the posterior mean given by the Gibbs sampler is
unbiased). Table \ref{tab:MSE} shows the estimated bias of the $100$
RMCMC system posterior means with respect to estimate given by the
Kalman filter. Table \ref{tab:SDs} shows the standard deviation of the
$100$ RMCMC system posterior means. We see that the standard deviation
(the accuracy $A$) is controlled below the imposed threshold of
$\beta_2=0.0125$. The lower row of Fig \ref{fig:kf-plots} are Q-Q
plots of the Kalman filter estimate and the weighted RMCMC samples
posterior distribution at the $1\%,2\%,\dots,99\%$ quantile from $1$
of the $100$ runs. The Q-Q plots for other components of and RMCMC
runs are similar to those presented. These Q-Q plots indicate that the
two distributions are roughly similar.

Comparison of the two distributions is difficult as the RMCMC samples
are not only weighted but are also dependent. Thus tests, such as the
Kolmogorov-Smirnov test \citep[e.g.\ see p.\ 35][]{opac-b1079143},
cannot be applied. 
\begin{table}[tbp]
\centering
 \caption{Tables of estimated bias of the $100$ system posterior means with respect to Kalman filter posterior mean.}
   \label{tab:MSE}
 \setlength{\tabcolsep}{2.5pt}
  \begin{tabular}{rddd}
     \toprule
     & \multicolumn{3}{c}{Last batch revealed}\\
     \cmidrule(r){2-4}
     & \multicolumn{1}{c}{$\widetilde{Y}_{1,6}$} &   \multicolumn{1}{c}{$\widetilde{Y}_{15,6}$} & \multicolumn{1}{c}{$\widetilde{Y}_{37,6}$}\\ 
   \midrule
   $X_{5,6}$  & 0.0009 & 0.0004 & 0.0004\\
   $X_{18,6}$ & -0.0012 & -0.0001 & 0.0049\\
   \bottomrule
  \end{tabular}%
  \hspace{13pt}
  \begin{tabular}{rddd}
     \toprule
     & \multicolumn{3}{c}{Last batch revealed}\\
     \cmidrule(r){2-4}
     & \multicolumn{1}{c}{$\widetilde{Y}_{1,6}$} &   \multicolumn{1}{c}{$\widetilde{Y}_{15,6}$} & \multicolumn{1}{c}{$\widetilde{Y}_{37,6}$}\\ 
   \midrule
   $X_{5,6}$  & 0.0009 & 0.0004 & 0.0004\\
$X_{18,6}$ & -0.0012 & -0.0001 & 0.0049\\
   \bottomrule
  \end{tabular}
\end{table}

\begin{table}[tbp]
  \centering
  \caption{Tables of the empirical standard deviation of posterior mean given by $100$ runs of the system.}\label{tab:SDs}
 \setlength{\tabcolsep}{2.5pt}
  \begin{tabular}{rddd}
    \toprule
    & \multicolumn{3}{c}{Last batch revealed}\\
    \cmidrule(r){2-4}
    &  \multicolumn{1}{c}{$\widetilde{Y}_{1,6}$} &   \multicolumn{1}{c}{$\widetilde{Y}_{15,6}$} & \multicolumn{1}{c}{$\widetilde{Y}_{37,6}$}\\
    \midrule
    $X_{5,6}$   & 0.0109 & 0.0023 & 0.0016\\
     $X_{18,6}$ & 0.0110 & 0.0019 & 0.0030 \\
     \bottomrule
  \end{tabular}%
  \hspace{13pt}
  \begin{tabular}{rddd}
    \toprule
    & \multicolumn{3}{c}{Last batch revealed}\\
    \cmidrule(r){2-4}
  &  \multicolumn{1}{c}{$\widetilde{Y}_{3,7}$} &   \multicolumn{1}{c}{$\widetilde{Y}_{10,7}$} &  \multicolumn{1}{c}{$\widetilde{Y}_{20,7}$}\\
    \midrule
    $X_{5,7}$   & 0.0065 & 0.0030 & 0.0026\\
    $X_{18,7}$   & 0.0077 & 0.0037 & 0.0022\\
     \bottomrule
  \end{tabular}
\end{table}

\section{Conclusion}\label{sec:conclusion}
We have presented a new method that produces estimates from a sequence
of distributions that maintains the accuracy at a user-specified
level. In \S \ref{sec:football} we demonstrated that the system is not
resumed each time an observation is revealed thus the samples are
reused. Therefore, we proceed with importance sampling whenever
possible. Further we attempt to reduce the size of the sample database
whenever possible (\S \ref{sec:control}), thus limiting the
computational effort of the update process and calculation of the
estimates or predictions. In \S \ref{sec:kalman_compare} we used a
linear Gaussian model to show that the system produced comparable
estimates to those given by the Kalman filter. Proving exactness of
the estimates produced by the system, if possible, is a topic for
future work.

For our system, we advocate using a standard MCMC method such as a
Metropolis-Hastings algorithm before resorting to another more
complicated method such as particle MCMC methods \citep{PMCMC} or
SMC$^2$ \citep{RSSB:RSSB1046}. By starting with a standard MCMC
approach, we avoid choosing the number of particles, choosing the
transition densities and the resampling step that comes with using a
particle filter, not to mention the higher computational cost.

\appendix
\section{Further System Results}\label{sec:more results}
\begin{figure}[H]
  \centering
  \subfloat[]{\label{fig:single_A}\includegraphics[width=0.33\textwidth]{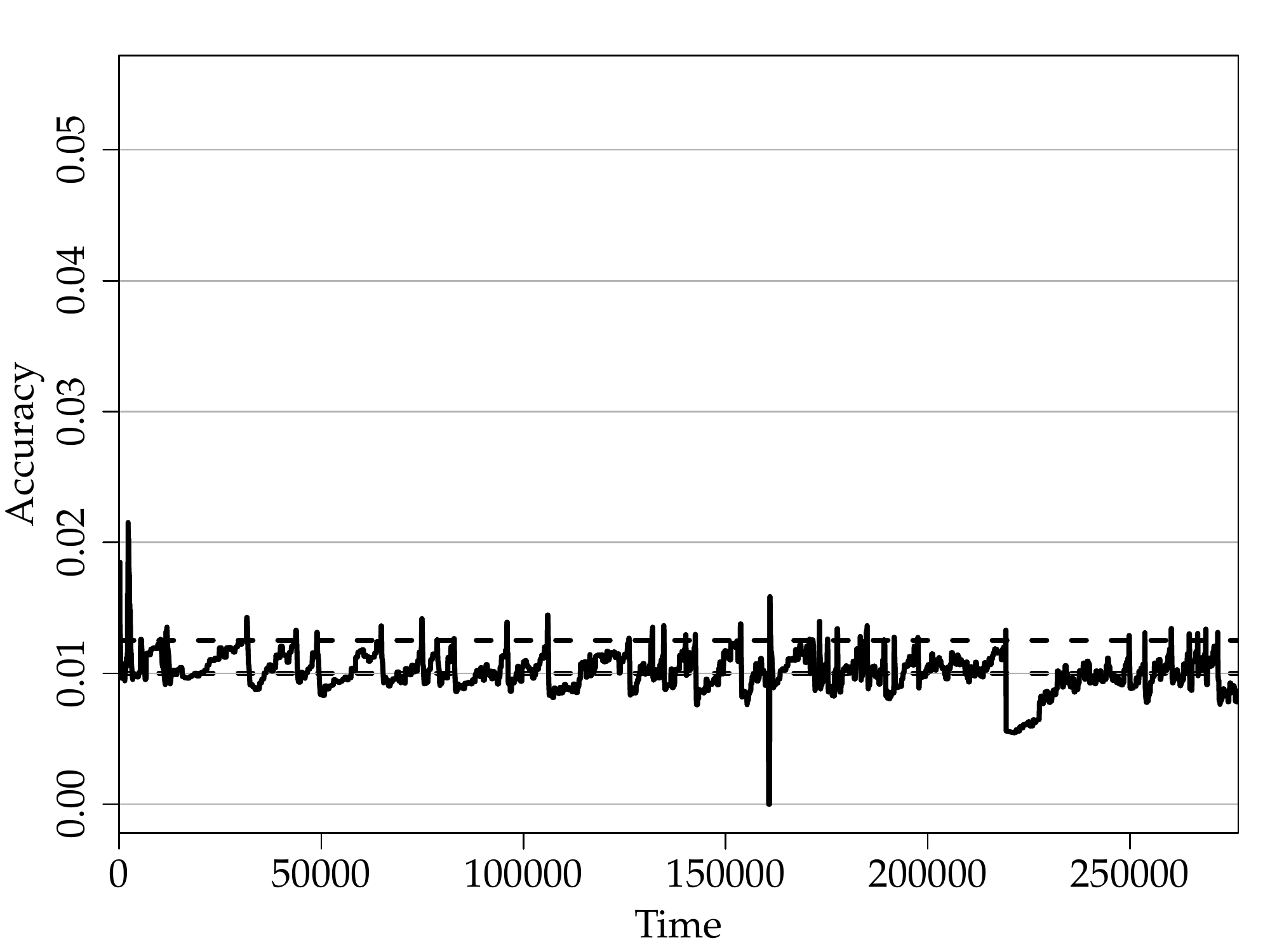}}
  \subfloat[]{\label{fig:single_Q}\includegraphics[width=0.33\textwidth]{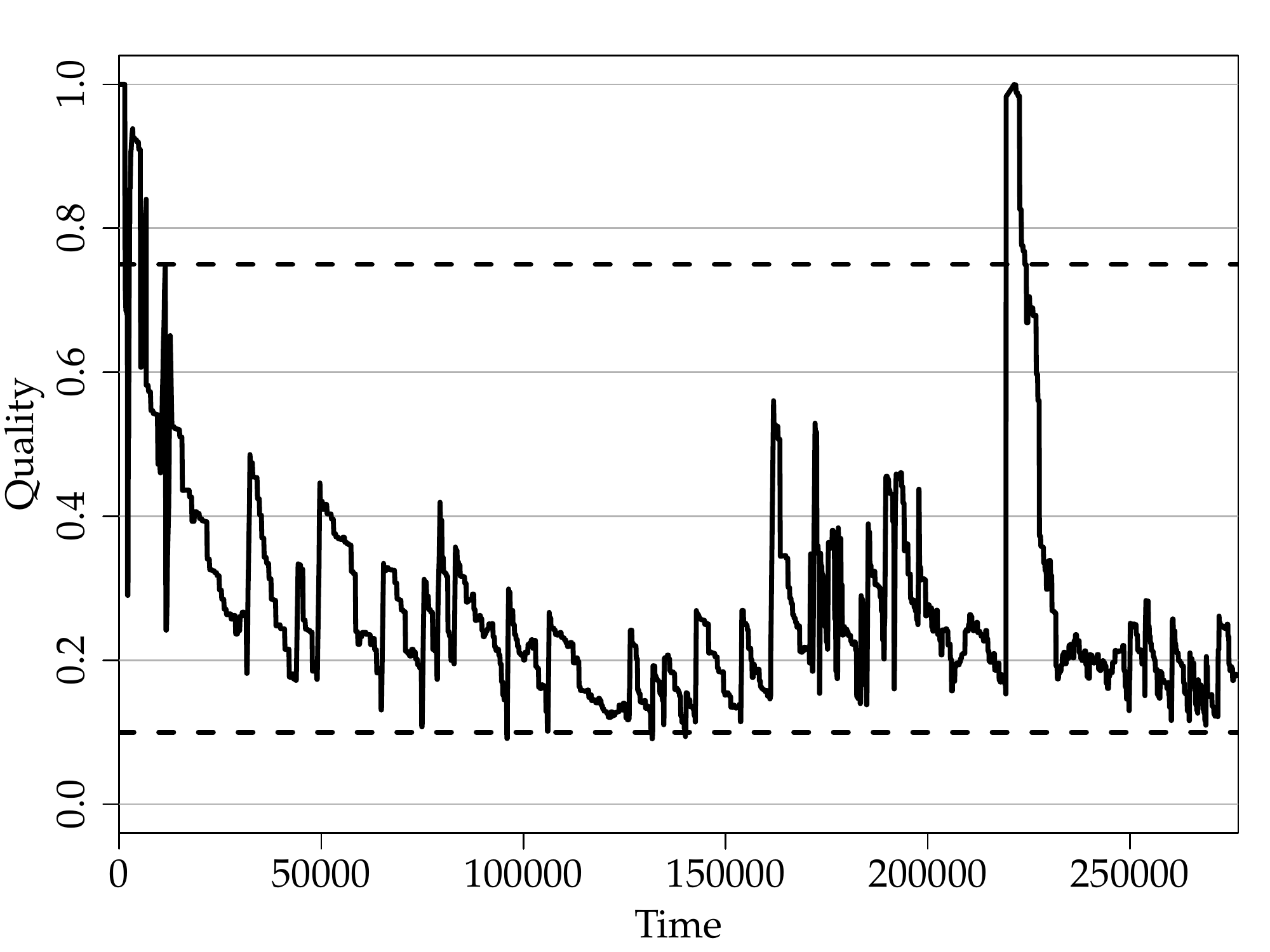}}
  \subfloat[]{\label{fig:single_N}\includegraphics[width=0.33\textwidth]{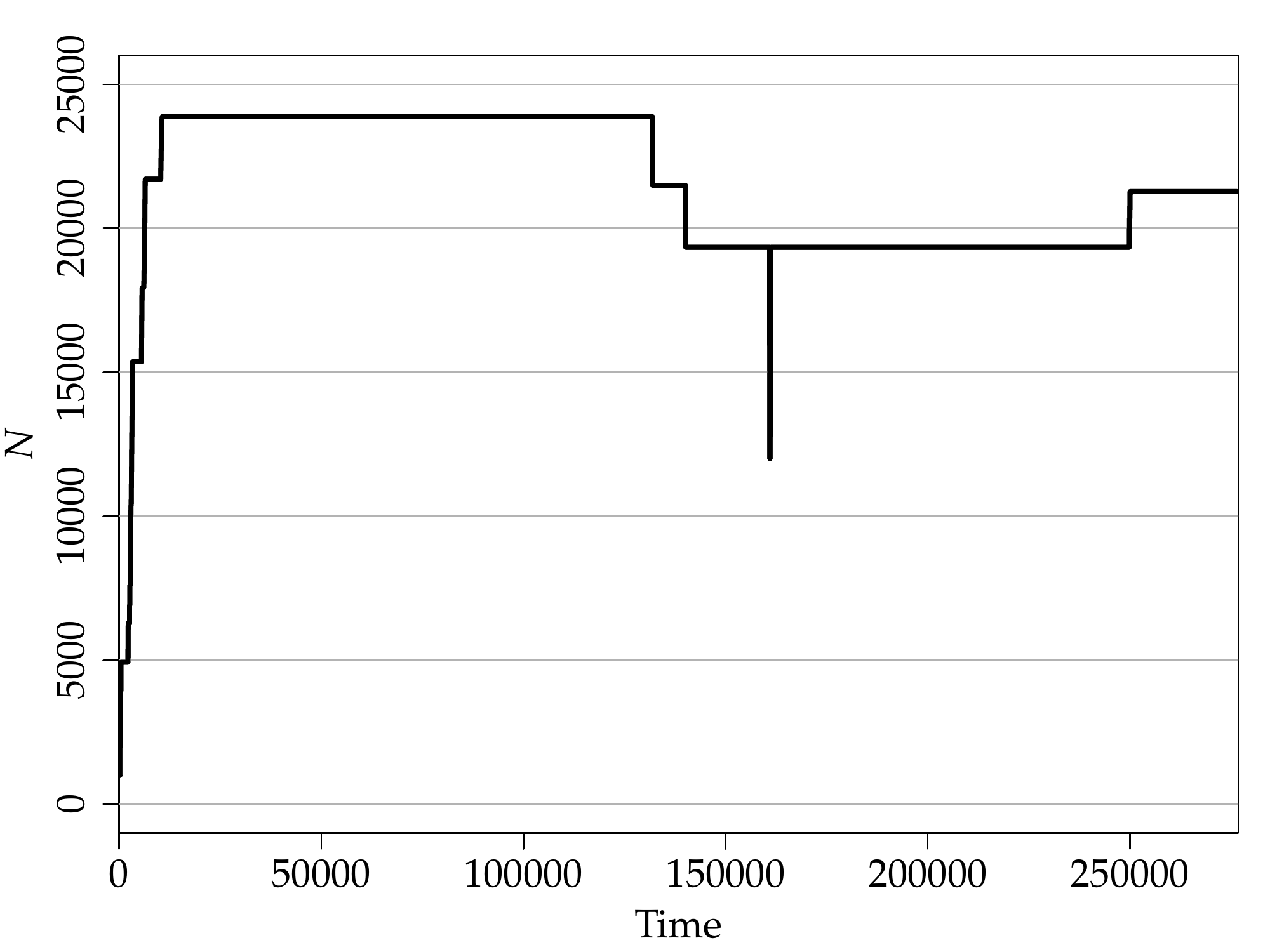}}

  \subfloat[]{\label{fig:weekly_A}\includegraphics[width=0.33\textwidth]{weekly_A.pdf}}
  \subfloat[]{\label{fig:weekly_Q}\includegraphics[width=0.33\textwidth]{weekly_Q.pdf}}
  \subfloat[]{\label{fig:weekly_N}\includegraphics[width=0.33\textwidth]{weekly_N.pdf}}

  \subfloat[]{\label{fig:monthly_A}\includegraphics[width=0.33\textwidth]{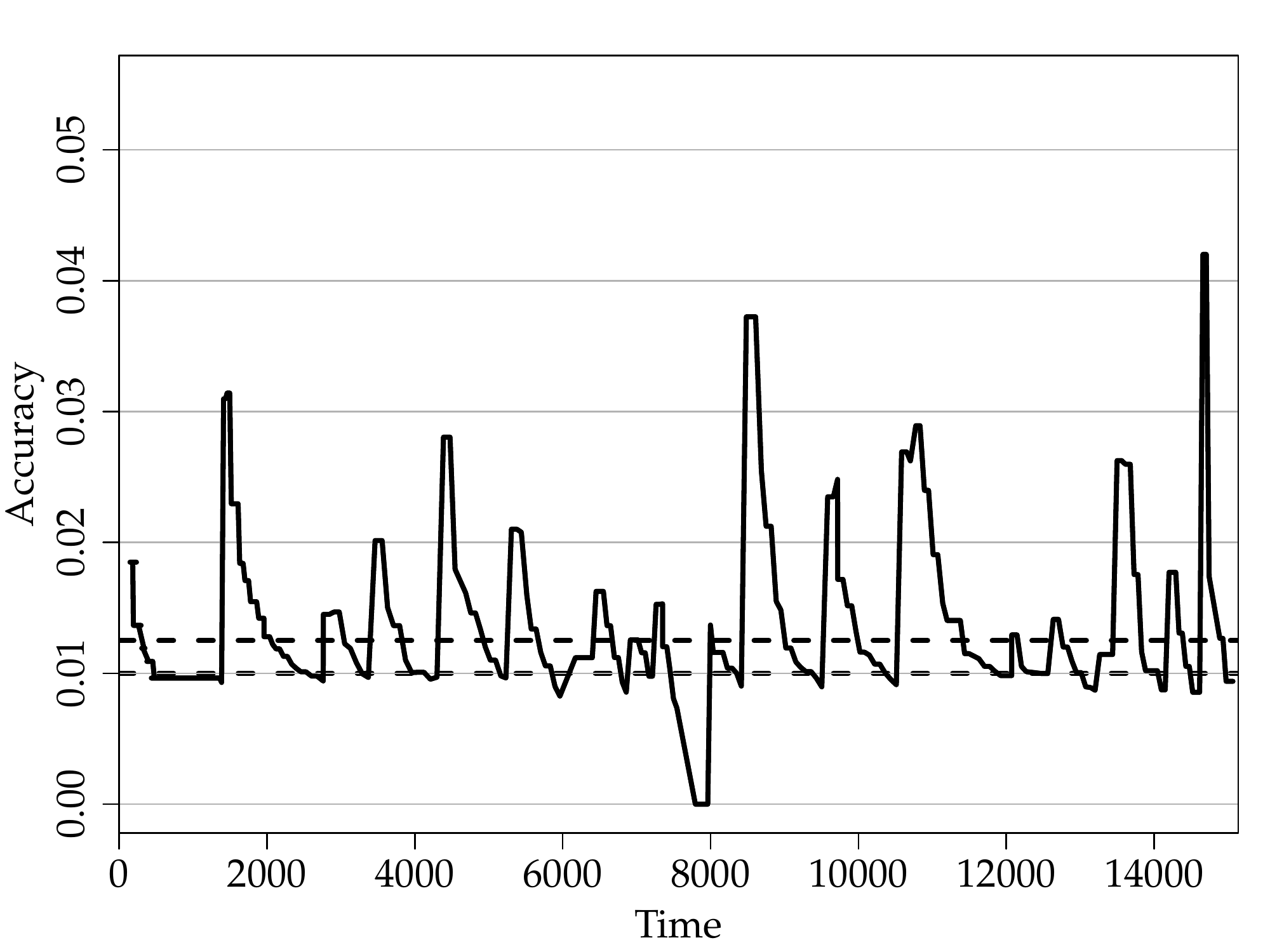}}
  \subfloat[]{\label{fig:monthly_Q}\includegraphics[width=0.33\textwidth]{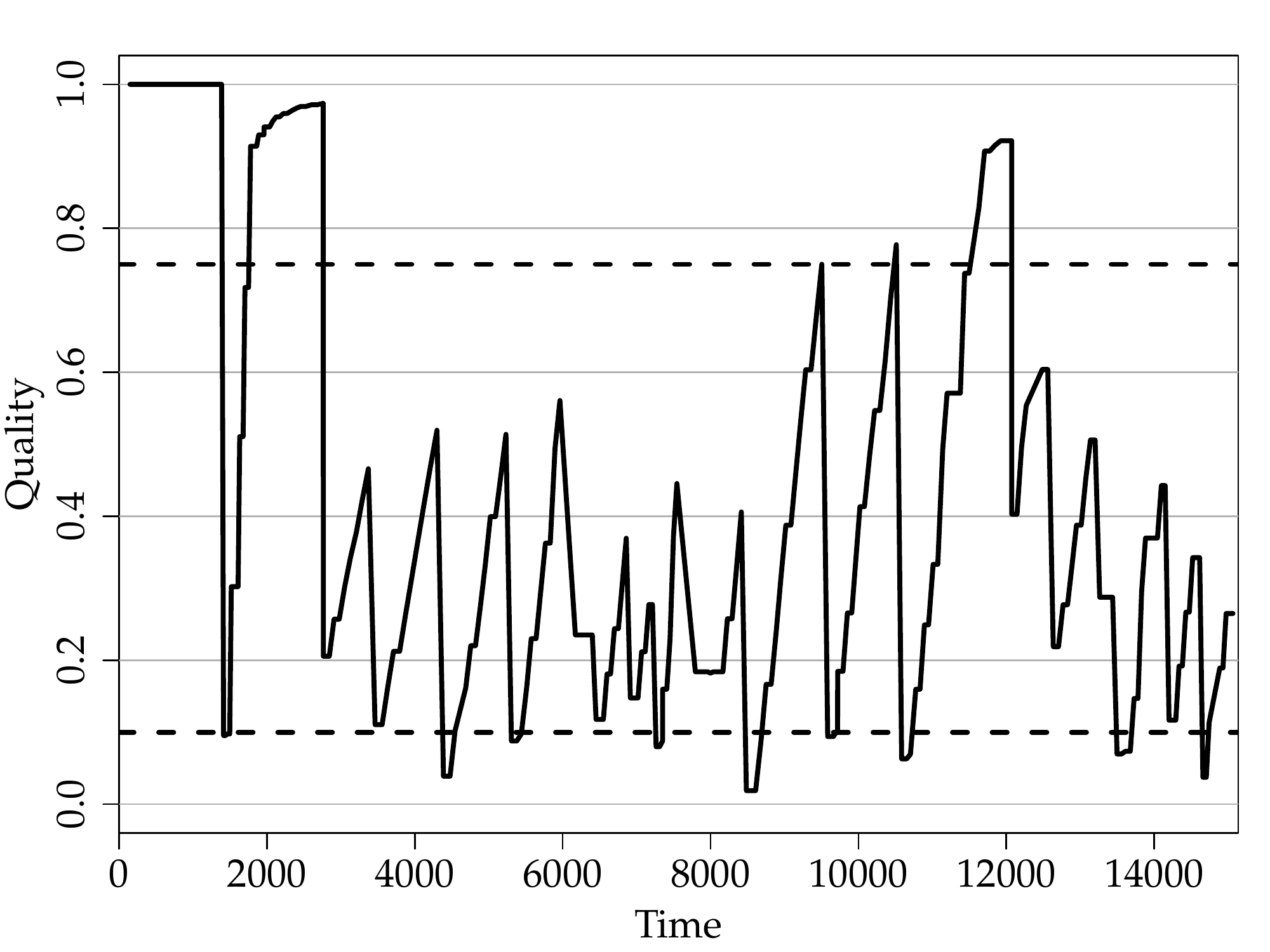}}
  \subfloat[]{\label{fig:monthly_N}\includegraphics[width=0.33\textwidth]{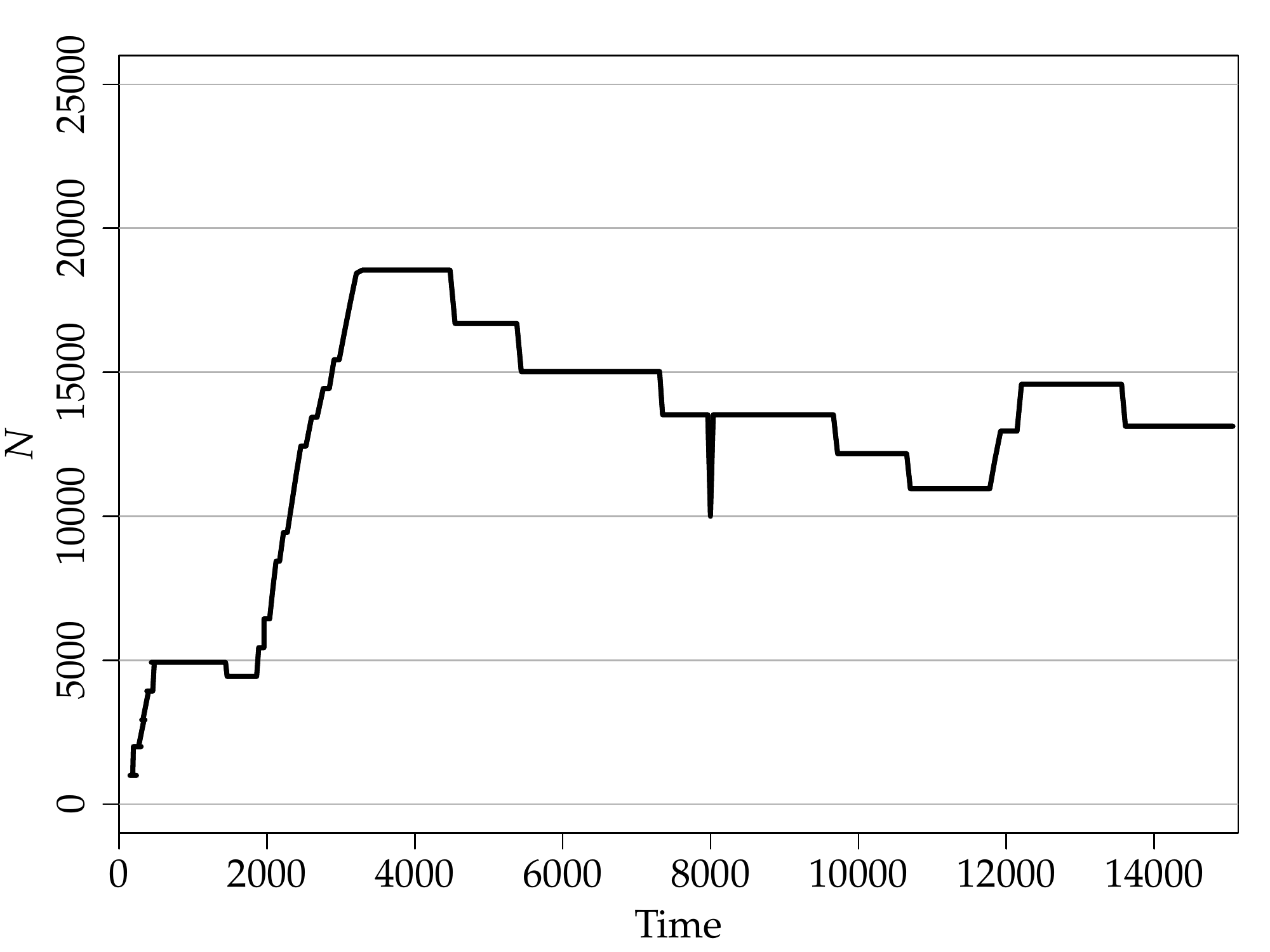}}

  \caption{Panel of plots of the system variables as new data, of
    varying size, are observed. Columns, from left to right, are the
    accuracy of the predictions ($A$), the quality of the samples
    ($Q$) and the Number of samples in the database $N$. Rows, from
    top to bottom are for individual, weekly and monthly data sizes.}
  \label{fig:system_parameters}
\end{figure}

In this section we present further results from \S 4.5 in the main
article. In Figure \ref{fig:system_parameters} we display the change
of the accuracy of the predictions ($A$), the quality of the samples
($Q$) and the number of samples in the database ($N$) as new data are
revealed. As expected, the larger the batch size the more frequently
the accuracy of the predictions and the quality of the samples exceed
the thresholds. Table \ref{tab:pred_single_ranks},
\ref{tab:pred_weekly_ranks} and \ref{tab:pred_monthly_ranks} present
the predicted end of 2012/13 English Premier league ranks for the
various batch sizes. The predictions are similar for all batch
sizes. This is unsurprising since the predictions are based on the
same data. Each probability (percentage) in Table
\ref{tab:pred_single_ranks}, \ref{tab:pred_weekly_ranks} and
\ref{tab:pred_monthly_ranks} are being controlled. More precisely, for
each team and each rank the standard deviation of the reported
probability (percentage) is being controlled below $\beta_2=0.0125$
(as set in the simulation in the main article \S 4.4).

\begin{figure}[H]
  \centering
  \subfloat[]{\label{fig:single_survival}\includegraphics[width=0.33\textwidth]{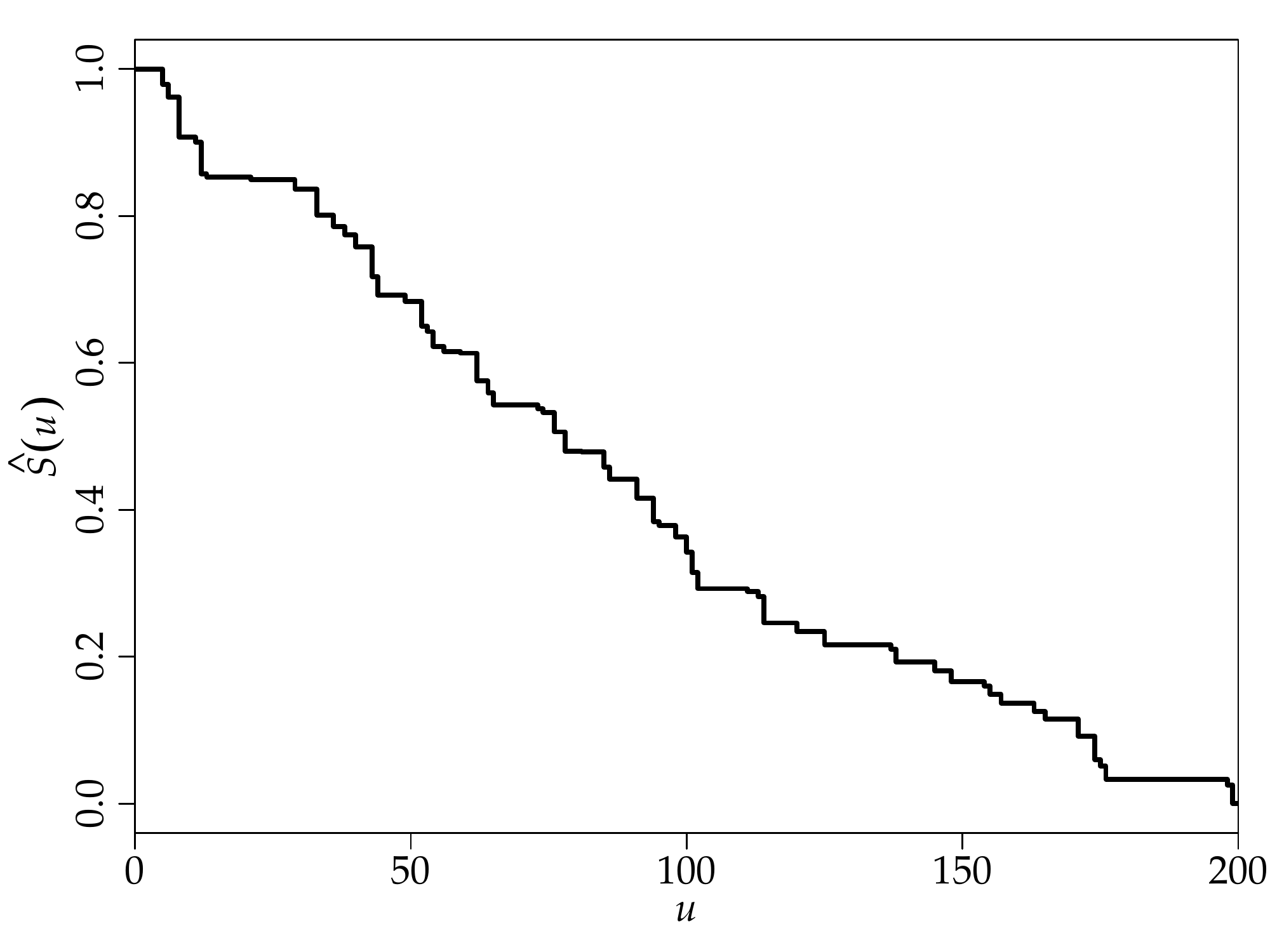}}
  \subfloat[]{\label{fig:weekly_survival}\includegraphics[width=0.33\textwidth]{KM-weekly.pdf}}
  \subfloat[]{\label{fig:monthly_survival}\includegraphics[width=0.33\textwidth]{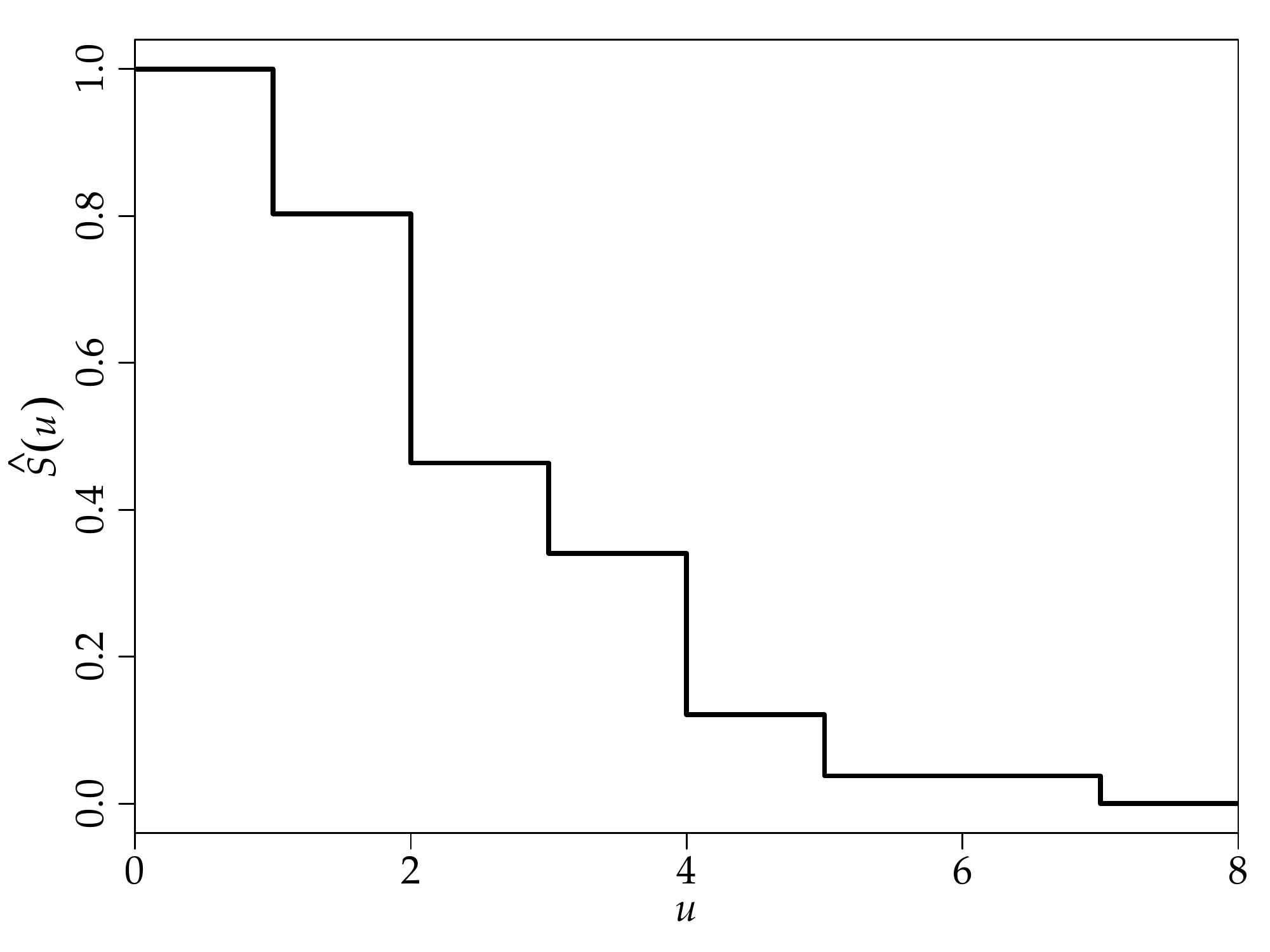}}
  \caption{Kaplan-Meier estimators of the survival of the samples as new data are
    observed using (a) individual results, (b) 7 day batches and (c)
    30 day batches.}
  \label{fig:system_survival}
\end{figure}

The survival of the samples as new data are observed varies greatly
depending on the batch size (Figure \ref{fig:system_survival}). From
the Kaplan-Meier estimators in Fig \ref{fig:single_survival},
\ref{fig:weekly_survival} and \ref{fig:monthly_survival} we observe
that smaller batches increases the number of batches a sample
survives. Hence, using smaller data batches results in samples being
reused more.
\begin{table}[H]
  \caption{Predicted end of season 2012/13 ranks for the English Premier League using individual results reported in percent.} 
  \label{tab:pred_single_ranks}
\centering
\tabcolsep=0.16cm
\centering
\begin{tabular}{lllllllllllllllllllll}
  \toprule
& \multicolumn{20}{c}{Rank} \\
    \cmidrule(r){2-21}
Team & 1 & 2 & 3 & 4 & 5 & 6 & 7 & 8 & 9 & 10 & 11 & 12 & 13 & 14 & 15 & 16 & 17 & 18 & 19 & 20 \\ 
  \midrule
Arsenal & 8 & 14 & 17 & 17 & 15 & 10 & 7 & 4 & 3 & 2 & 1 & 1 & 1 & 0 & 0 & 0 & 0 & 0 & 0 & 0 \\ 
  Aston Villa & 0 & 0 & 1 & 1 & 3 & 3 & 5 & 6 & 7 & 7 & 8 & 8 & 8 & 7 & 8 & 7 & 6 & 6 & 5 & 4 \\ 
  Chelsea & 9 & 15 & 19 & 17 & 13 & 9 & 6 & 4 & 2 & 2 & 1 & 1 & 1 & 0 & 0 & 0 & 0 & 0 & 0 & 0 \\ 
  Everton & 1 & 2 & 6 & 8 & 11 & 11 & 12 & 10 & 9 & 7 & 6 & 5 & 4 & 3 & 2 & 1 & 2 & 1 & 1 & 0 \\ 
  Fulham & 0 & 1 & 2 & 4 & 5 & 7 & 8 & 9 & 9 & 8 & 8 & 6 & 6 & 7 & 5 & 4 & 4 & 3 & 2 & 1 \\ 
  Liverpool & 2 & 5 & 8 & 11 & 13 & 13 & 10 & 9 & 7 & 6 & 5 & 3 & 2 & 2 & 2 & 1 & 1 & 0 & 0 & 0 \\ 
  Man City & 32 & 28 & 18 & 10 & 5 & 3 & 2 & 1 & 0 & 0 & 0 & 0 & 0 & 0 & 0 & 0 & 0 & 0 & 0 & 0 \\ 
  Man United & 46 & 26 & 12 & 7 & 4 & 2 & 1 & 1 & 0 & 0 & 0 & 0 & 0 & 0 & 0 & 0 & 0 & 0 & 0 & 0 \\ 
  Newcastle & 0 & 1 & 2 & 3 & 5 & 6 & 8 & 9 & 10 & 8 & 8 & 7 & 6 & 6 & 5 & 5 & 4 & 3 & 2 & 2 \\ 
  Norwich & 0 & 0 & 0 & 1 & 1 & 2 & 3 & 4 & 4 & 6 & 6 & 8 & 8 & 8 & 8 & 9 & 9 & 8 & 9 & 7 \\ 
  QPR & 0 & 0 & 0 & 0 & 0 & 1 & 2 & 2 & 3 & 5 & 5 & 6 & 6 & 7 & 8 & 9 & 9 & 11 & 13 & 12 \\ 
  Reading & 0 & 0 & 0 & 0 & 1 & 1 & 2 & 2 & 4 & 4 & 5 & 5 & 7 & 7 & 7 & 9 & 9 & 10 & 12 & 14 \\ 
  Southampton & 0 & 0 & 0 & 1 & 1 & 2 & 2 & 3 & 3 & 4 & 5 & 5 & 6 & 7 & 8 & 8 & 10 & 11 & 11 & 13 \\ 
  Stoke & 0 & 0 & 1 & 1 & 2 & 3 & 4 & 6 & 7 & 7 & 7 & 8 & 8 & 7 & 8 & 7 & 7 & 7 & 7 & 5 \\ 
  Sunderland & 0 & 0 & 1 & 2 & 4 & 5 & 6 & 8 & 8 & 8 & 8 & 8 & 7 & 7 & 6 & 6 & 5 & 4 & 4 & 3 \\ 
  Swansea & 0 & 0 & 0 & 1 & 1 & 3 & 4 & 5 & 6 & 6 & 7 & 8 & 8 & 8 & 8 & 8 & 8 & 7 & 7 & 6 \\ 
  Tottenham & 2 & 7 & 12 & 14 & 14 & 12 & 10 & 8 & 6 & 4 & 3 & 2 & 2 & 2 & 1 & 1 & 0 & 0 & 0 & 0 \\ 
  West Brom & 0 & 0 & 0 & 1 & 2 & 3 & 4 & 5 & 6 & 8 & 7 & 8 & 8 & 8 & 7 & 8 & 7 & 7 & 5 & 5 \\ 
  West Ham & 0 & 0 & 0 & 0 & 1 & 1 & 2 & 3 & 3 & 4 & 5 & 5 & 6 & 7 & 8 & 9 & 10 & 11 & 11 & 14 \\ 
  Wigan & 0 & 0 & 0 & 0 & 1 & 1 & 2 & 2 & 3 & 4 & 5 & 6 & 7 & 8 & 9 & 9 & 10 & 10 & 11 & 13 \\ 
   \bottomrule
\end{tabular}
\end{table}
\begin{table}[H]
  \caption{Predicted end of season 2012/13 ranks for the English Premier League using 7 day batches reported in percent.}
  \label{tab:pred_weekly_ranks}
\centering
\tabcolsep=0.16cm
\begin{tabular}{lllllllllllllllllllll}
  \toprule
& \multicolumn{20}{c}{Rank} \\
   \cmidrule(r){2-21}
   Team & 1 & 2 & 3 & 4 & 5 & 6 & 7 & 8 & 9 & 10 & 11 & 12 & 13 & 14 & 15 & 16 & 17 & 18 & 19 & 20 \\ 
  \midrule
Arsenal & 8 & 14 & 17 & 19 & 13 & 9 & 6 & 4 & 3 & 2 & 2 & 1 & 1 & 0 & 0 & 0 & 0 & 0 & 0 & 0 \\ 
  Aston Villa & 0 & 0 & 1 & 1 & 3 & 4 & 5 & 6 & 7 & 7 & 7 & 9 & 8 & 8 & 7 & 7 & 6 & 6 & 5 & 4 \\ 
  Chelsea & 9 & 15 & 21 & 16 & 12 & 9 & 6 & 4 & 3 & 1 & 1 & 1 & 1 & 0 & 0 & 0 & 0 & 0 & 0 & 0 \\ 
  Everton & 1 & 3 & 5 & 8 & 11 & 12 & 13 & 10 & 8 & 6 & 5 & 4 & 3 & 3 & 2 & 2 & 1 & 1 & 1 & 0 \\ 
  Fulham & 0 & 1 & 2 & 3 & 5 & 6 & 8 & 9 & 9 & 9 & 9 & 7 & 7 & 5 & 5 & 4 & 4 & 3 & 3 & 2 \\ 
  Liverpool & 2 & 4 & 7 & 11 & 15 & 13 & 10 & 9 & 7 & 6 & 4 & 3 & 3 & 2 & 2 & 1 & 1 & 1 & 0 & 0 \\ 
  Man City & 29 & 27 & 18 & 10 & 6 & 4 & 2 & 1 & 0 & 0 & 0 & 0 & 0 & 0 & 0 & 0 & 0 & 0 & 0 & 0 \\ 
  Man United & 47 & 26 & 14 & 7 & 3 & 1 & 1 & 0 & 0 & 0 & 0 & 0 & 0 & 0 & 0 & 0 & 0 & 0 & 0 & 0 \\ 
  Newcastle & 0 & 1 & 2 & 3 & 5 & 7 & 9 & 9 & 8 & 9 & 8 & 7 & 7 & 6 & 6 & 4 & 3 & 3 & 2 & 1 \\ 
  Norwich & 0 & 0 & 0 & 0 & 2 & 2 & 3 & 4 & 5 & 6 & 6 & 8 & 7 & 8 & 7 & 9 & 9 & 9 & 9 & 8 \\ 
  QPR & 0 & 0 & 0 & 0 & 1 & 1 & 2 & 2 & 3 & 4 & 5 & 6 & 6 & 7 & 7 & 8 & 9 & 12 & 13 & 13 \\ 
  Reading & 0 & 0 & 0 & 0 & 1 & 2 & 2 & 2 & 3 & 4 & 5 & 6 & 5 & 7 & 8 & 8 & 10 & 11 & 11 & 15 \\ 
  Southampton & 0 & 0 & 0 & 0 & 0 & 1 & 2 & 2 & 3 & 5 & 5 & 5 & 6 & 7 & 8 & 9 & 9 & 10 & 12 & 13 \\ 
  Stoke & 0 & 0 & 0 & 1 & 2 & 3 & 4 & 6 & 6 & 7 & 8 & 8 & 8 & 7 & 7 & 8 & 7 & 7 & 7 & 5 \\ 
  Sunderland & 0 & 0 & 1 & 3 & 3 & 5 & 7 & 8 & 8 & 8 & 8 & 8 & 8 & 7 & 7 & 5 & 5 & 4 & 3 & 2 \\ 
  Swansea & 0 & 0 & 0 & 1 & 1 & 2 & 3 & 4 & 6 & 6 & 7 & 7 & 8 & 8 & 8 & 8 & 9 & 8 & 7 & 6 \\ 
  Tottenham & 4 & 8 & 11 & 14 & 14 & 13 & 10 & 8 & 6 & 4 & 3 & 2 & 1 & 1 & 1 & 1 & 0 & 0 & 0 & 0 \\ 
  West Brom & 0 & 0 & 0 & 1 & 2 & 3 & 4 & 6 & 7 & 7 & 7 & 7 & 8 & 7 & 8 & 8 & 7 & 6 & 5 & 5 \\ 
  West Ham & 0 & 0 & 0 & 0 & 1 & 1 & 2 & 3 & 3 & 5 & 4 & 5 & 7 & 7 & 9 & 9 & 9 & 11 & 11 & 14 \\ 
  Wigan & 0 & 0 & 0 & 0 & 1 & 1 & 2 & 3 & 3 & 5 & 5 & 6 & 7 & 9 & 8 & 9 & 10 & 10 & 11 & 12 \\ 
   \bottomrule
\end{tabular}
\end{table}
\begin{table}[H]
  \caption{Predicted end of season 2012/13 ranks for the English Premier League using 30 day batches reported in percent.}
  \label{tab:pred_monthly_ranks}
\centering
\tabcolsep=0.16cm
\centering
\begin{tabular}{lllllllllllllllllllll}
  \toprule
& \multicolumn{20}{c}{Rank} \\
   \cmidrule(r){2-21}
Team & 1 & 2 & 3 & 4 & 5 & 6 & 7 & 8 & 9 & 10 & 11 & 12 & 13 & 14 & 15 & 16 & 17 & 18 & 19 & 20 \\ 
  \midrule
Arsenal & 8 & 15 & 19 & 16 & 12 & 10 & 6 & 5 & 3 & 2 & 2 & 1 & 1 & 0 & 0 & 0 & 0 & 0 & 0 & 0 \\ 
  Aston Villa & 0 & 0 & 1 & 1 & 2 & 4 & 6 & 6 & 7 & 7 & 7 & 8 & 8 & 7 & 6 & 7 & 6 & 6 & 5 & 3 \\ 
  Chelsea & 10 & 16 & 20 & 16 & 12 & 9 & 6 & 4 & 2 & 2 & 2 & 1 & 0 & 1 & 0 & 0 & 0 & 0 & 0 & 0 \\ 
  Everton & 1 & 2 & 5 & 8 & 11 & 10 & 10 & 10 & 9 & 8 & 5 & 5 & 3 & 3 & 2 & 1 & 2 & 1 & 1 & 0 \\ 
  Fulham & 0 & 1 & 2 & 4 & 5 & 8 & 9 & 9 & 8 & 9 & 8 & 8 & 6 & 5 & 6 & 4 & 3 & 3 & 2 & 1 \\ 
  Liverpool & 2 & 4 & 7 & 11 & 13 & 12 & 11 & 9 & 7 & 6 & 4 & 4 & 3 & 2 & 2 & 1 & 1 & 1 & 0 & 0 \\ 
  Man City & 29 & 28 & 17 & 11 & 6 & 3 & 2 & 1 & 1 & 0 & 0 & 0 & 0 & 0 & 0 & 0 & 0 & 0 & 0 & 0 \\ 
  Man United & 47 & 24 & 14 & 7 & 4 & 2 & 1 & 0 & 0 & 0 & 0 & 0 & 0 & 0 & 0 & 0 & 0 & 0 & 0 & 0 \\ 
  Newcastle & 0 & 1 & 2 & 4 & 5 & 6 & 9 & 9 & 9 & 9 & 9 & 7 & 7 & 6 & 4 & 4 & 3 & 3 & 2 & 1 \\ 
  Norwich & 0 & 0 & 0 & 0 & 1 & 2 & 2 & 4 & 5 & 6 & 7 & 7 & 7 & 8 & 8 & 8 & 8 & 8 & 9 & 9 \\ 
  QPR & 0 & 0 & 0 & 0 & 0 & 1 & 2 & 2 & 3 & 4 & 4 & 4 & 7 & 7 & 8 & 9 & 10 & 11 & 12 & 14 \\ 
  Reading & 0 & 0 & 0 & 0 & 1 & 2 & 2 & 2 & 3 & 5 & 4 & 5 & 7 & 7 & 8 & 8 & 9 & 10 & 12 & 14 \\ 
  Southampton & 0 & 0 & 0 & 1 & 1 & 2 & 2 & 2 & 3 & 4 & 6 & 6 & 6 & 7 & 9 & 8 & 10 & 10 & 10 & 13 \\ 
  Stoke & 0 & 0 & 0 & 1 & 2 & 2 & 4 & 6 & 5 & 7 & 7 & 8 & 7 & 8 & 8 & 8 & 8 & 7 & 6 & 5 \\ 
  Sunderland & 0 & 0 & 1 & 2 & 4 & 6 & 6 & 7 & 8 & 8 & 7 & 8 & 8 & 7 & 6 & 6 & 4 & 5 & 4 & 2 \\ 
  Swansea & 0 & 0 & 0 & 1 & 2 & 3 & 4 & 5 & 7 & 6 & 6 & 7 & 8 & 9 & 8 & 9 & 7 & 7 & 7 & 5 \\ 
  Tottenham & 3 & 7 & 10 & 14 & 15 & 13 & 10 & 7 & 6 & 4 & 3 & 2 & 2 & 1 & 1 & 1 & 0 & 0 & 0 & 0 \\ 
  West Brom & 0 & 0 & 1 & 1 & 2 & 4 & 4 & 5 & 7 & 8 & 8 & 7 & 8 & 7 & 7 & 7 & 6 & 7 & 6 & 5 \\ 
  West Ham & 0 & 0 & 0 & 0 & 1 & 1 & 2 & 3 & 3 & 4 & 4 & 5 & 6 & 7 & 8 & 9 & 10 & 10 & 12 & 13 \\ 
  Wigan & 0 & 0 & 0 & 0 & 1 & 1 & 2 & 2 & 3 & 4 & 5 & 6 & 6 & 7 & 7 & 9 & 11 & 11 & 12 & 13 \\ 
   \bottomrule
\end{tabular}
\end{table}


\begin{thebibliography}{24}
\providecommand{\natexlab}[1]{#1}
\providecommand{\url}[1]{\texttt{#1}}
\expandafter\ifx\csname urlstyle\endcsname\relax
  \providecommand{\doi}[1]{doi: #1}\else
  \providecommand{\doi}{doi: \begingroup \urlstyle{rm}\Url}\fi

\bibitem[Andrieu et~al.(2010)Andrieu, Doucet, and Holenstein]{PMCMC}
C.~Andrieu, A.~Doucet, and R.~Holenstein.
\newblock Particle {M}arkov chain {M}onte {C}arlo methods.
\newblock \emph{Journal of the Royal Statistical Society: Series B},
  72\penalty0 (3):\penalty0 269--342, 2010.

\bibitem[Brooks et~al.(2011)Brooks, Gelman, Jones, and
  Meng]{MonteCarloHandbook}
S.~Brooks, A.~Gelman, G.~Jones, and X.~Meng.
\newblock \emph{Handbook of Markov Chain Monte Carlo}.
\newblock CRC Press, 2011.

\bibitem[Chopin(2002)]{chopin2002sequential}
N.~Chopin.
\newblock A sequential particle filter method for static models.
\newblock \emph{Biometrika}, 89\penalty0 (3):\penalty0 539--552, 2002.

\bibitem[Chopin et~al.(2013)Chopin, Jacob, and Papaspiliopoulos]{RSSB:RSSB1046}
N.~Chopin, P.~E. Jacob, and O.~Papaspiliopoulos.
\newblock {SMC2}: an efficient algorithm for sequential analysis of state space
  models.
\newblock \emph{Journal of the Royal Statistical Society: Series B},
  75\penalty0 (3):\penalty0 397--426, 2013.

\bibitem[Del~Moral et~al.(2006)Del~Moral, Doucet, and Jasra]{del2006sequential}
P.~Del~Moral, A.~Doucet, and A.~Jasra.
\newblock Sequential {M}onte {C}arlo samplers.
\newblock \emph{Journal of the Royal Statistical Society: Series B (Statistical
  Methodology)}, 68\penalty0 (3):\penalty0 411--436, 2006.

\bibitem[Dixon and Coles(1997)]{RSSC:RSSC065}
M.~J. Dixon and S.~G. Coles.
\newblock Modelling association football scores and inefficiencies in the
  football betting market.
\newblock \emph{Journal of the Royal Statistical Society: Series C},
  46\penalty0 (2):\penalty0 265--280, 1997.

\bibitem[Doucet et~al.(2001)Doucet, de~Freitas, and Gordon]{smcm}
A.~Doucet, N.~de~Freitas, and N.~Gordon, editors.
\newblock \emph{Sequential Monte Carlo Methods in Practice}.
\newblock Springer, 2001.

\bibitem[Flegal and Jones(2010)]{flegal_batch}
J.~M. Flegal and G.~L. Jones.
\newblock Batch means and spectral variance estimators in {M}arkov chain
  {M}onte {C}arlo.
\newblock \emph{The Annals of Statistics}, 38\penalty0 (2):\penalty0
  1034--1070, 2010.

\bibitem[Geman and Geman(1984)]{geman1984stochastic}
S.~Geman and D.~Geman.
\newblock Stochastic relaxation, {G}ibbs distributions, and the {B}ayesian
  restoration of images.
\newblock \emph{Pattern Analysis and Machine Intelligence, IEEE Transactions
  on}, \penalty0 (6):\penalty0 721--741, 1984.

\bibitem[Geyer(1992)]{practical_MCMC_geyer}
C.~J. Geyer.
\newblock Practical {M}arkov chain {M}onte {C}arlo.
\newblock \emph{Statistical Science}, 7\penalty0 (4):\penalty0 473--483, 1992.

\bibitem[Glickman and Stern(1998)]{doi:10.1080/01621459.1998.10474084}
M.~E. Glickman and H.~S. Stern.
\newblock A state-space model for {N}ational {F}ootball {L}eague scores.
\newblock \emph{Journal of the American Statistical Association}, 93\penalty0
  (441):\penalty0 25--35, 1998.

\bibitem[Gordon et~al.(1993)Gordon, Salmond, and Smith]{gordon}
N.~Gordon, D.~Salmond, and A.~Smith.
\newblock {Novel approach to nonlinear/non-Gaussian Bayesian state estimation}.
\newblock \emph{Radar and Signal Processing, IEE Proceedings F}, 140\penalty0
  (2):\penalty0 107 --113, 1993.

\bibitem[Gramacy et~al.(2010)Gramacy, Samworth, and King]{gramacy}
R.~Gramacy, R.~Samworth, and R.~King.
\newblock Importance tempering.
\newblock \emph{Statistics and Computing}, 20\penalty0 (1):\penalty0 1--7,
  2010.

\bibitem[Hastings(1970)]{HASTINGS01041970}
W.~K. Hastings.
\newblock Monte {C}arlo sampling methods using {M}arkov chains and their
  applications.
\newblock \emph{Biometrika}, 57\penalty0 (1):\penalty0 97--109, 1970.

\bibitem[Hesterberg(1995)]{hesterberg1995weighted}
T.~Hesterberg.
\newblock Weighted average importance sampling and defensive mixture
  distributions.
\newblock \emph{Technometrics}, 37\penalty0 (2):\penalty0 185--194, 1995.

\bibitem[Hintze and Nelson(1998)]{violin-plots}
J.~L. Hintze and R.~D. Nelson.
\newblock Violin plots: A box plot-density trace synergism.
\newblock \emph{The American Statistician}, 52\penalty0 (2):\penalty0 181--184,
  1998.

\bibitem[Jones(2004)]{jones2004markov}
G.~L. Jones.
\newblock On the {M}arkov chain central limit theorem.
\newblock \emph{Probability surveys}, 1:\penalty0 299--320, 2004.

\bibitem[Kalman(1960)]{kalman1960new}
R.~E. Kalman.
\newblock A new approach to linear filtering and prediction problems.
\newblock \emph{Journal of basic Engineering}, 82\penalty0 (1):\penalty0
  35--45, 1960.

\bibitem[Kaplan and Meier(1958)]{km}
E.~Kaplan and P.~Meier.
\newblock Nonparametric estimation from incomplete observations.
\newblock \emph{Journal of the American Statistical Association,}, 53:\penalty0
  457--481, 1958.

\bibitem[Kitagawa(2014)]{kitagawa2014computational}
G.~Kitagawa.
\newblock Computational aspects of sequential {M}onte {C}arlo filter and
  smoother.
\newblock \emph{Annals of the Institute of Statistical Mathematics}, pages
  1--29, 2014.

\bibitem[Lahiri(2003)]{lahiri_resampling}
S.~Lahiri.
\newblock \emph{Resampling Methods for Dependent Data}.
\newblock Springer, 2003.

\bibitem[Lehmann and D'Abrera(1975)]{opac-b1079143}
E.~L. Lehmann and H.~D'Abrera.
\newblock \emph{Nonparametrics: Statistical Methods Based on Ranks}.
\newblock Holden-Day, 1975.

\bibitem[Liu(2008)]{liu2008monte}
J.~S. Liu.
\newblock \emph{Monte {C}arlo {S}trategies in {S}cientific {C}omputing}.
\newblock Springer, 2008.

\bibitem[Metropolis et~al.(1953)Metropolis, Rosenbluth, Rosenbluth, Teller, and
  Teller]{annealing2}
N.~Metropolis, A.~W. Rosenbluth, M.~N. Rosenbluth, A.~H. Teller, and E.~Teller.
\newblock Equation of state calculations by fast computing machines.
\newblock \emph{The Journal of Chemical Physics}, 21\penalty0 (6):\penalty0
  1087--1092, 1953.

\end{thebibliography}

\end{document}